\documentclass[twocolumn]{aastex7}

\graphicspath{{./}{figures/}}
\usepackage{xcolor}
\usepackage{amsmath}
\usepackage{textcomp, gensymb}

\newcommand{\TESS}{\textit{TESS}}
\newcommand{\CHEOPS}{\textit{CHEOPS}}

\renewcommand{\Re}{R$_\oplus$}

\newcommand{\Rsun}{R$_\odot$}
\newcommand{\Lsun}{L$_\odot$}

\newcommand{\thisstar}{TIC~384984325}
\newcommand{\planetb}{TOI-6109\,b}
\newcommand{\perb}{$5.6904^{+0.0004}_{-0.0004}$}
\newcommand{\radb}{$4.87^{+0.16}_{-0.12}$}
\newcommand{\planetc}{TOI-6109\,c}
\newcommand{\perc}{$8.5388^{+0.0006}_{-0.0005}$}
\newcommand{\radc}{$4.83^{+0.07}_{-0.06}$}

\usepackage{etoolbox}
\makeatletter
\patchcmd{\ltx@foottext}{%
  .5\textwidth\advance\hsize-18pt}{%
  \linewidth\advance\hsize-1.8em%
}{}{}
\makeatother

\begin{document}

\title{THYME XIII: Two young Neptunes orbiting a 75-Myr star in the Alpha Persei Cluster}
\shorttitle{TOI-6109}

\author[0000-0002-1092-2995]{Anne Dattilo}
\affiliation{Department of Astronomy \& Astrophysics, 525 Davey Laboratory, 251 Pollock Road, Penn State, University Park, PA, 16802, USA}
\affiliation{Center for Exoplanets and Habitable Worlds, Penn State University, 525 Davey Laboratory, 251 Pollock Road, University Park, PA, 16802, USA}
\affiliation{Department of Astronomy \& Astrophysics, University of California Santa Cruz, 1156 High Street, Santa Cruz, CA, 95064, USA}
\email{adattilo@psu.edu}

\author[0000-0001-7246-5438]{Andrew M. Vanderburg}
\altaffiliation{Sloan Research Fellow}
\affiliation{Center for Astrophysics \textbar \ Harvard \& Smithsonian, 60 Garden Street, Cambridge, MA 02138, USA}
\email{avanderburg@cfa.harvard.edu}

\author[0000-0002-8399-472X]{Madyson G. Barber}
\altaffiliation{NSF Graduate Research Fellow}
\affiliation{Department of Physics and Astronomy, The University of North Carolina at Chapel Hill, Chapel Hill, NC 27599, USA}
\email{madysonb@live.unc.edu}

\author[0000-0003-3654-1602]{Andrew W. Mann}
\affiliation{Department of Physics and Astronomy, The University of North Carolina at Chapel Hill, Chapel Hill, NC 27599, USA}
\email{awmann@unc.edu}

\author[0000-0002-6549-9792]{Ronan Kerr}
\affiliation{Dunlap Institute for Astronomy \& Astrophysics, University of Toronto, Toronto, ON M5S 3H4, Canada}
\email{ronan.kerr@utoronto.ca}

\author[0000-0001-9811-568X]{Adam L. Kraus}
\affiliation{The University of Texas at Austin, 2515 Speedway Ave, Austin, TX 78712-1205, USA}
\email{alk@astro.as.utexas.edu}

\author[0000-0003-3888-3753]{Joseph R. Livesey}
\affiliation{Department of Astronomy, University of Wisconsin–Madison, 475 N Charter St, Madison, WI 53706, USA}
\email{jrlivesey@wisc.edu}

\author[0000-0001-8621-6731]{Cristilyn Watkins}
\affiliation{Center for Astrophysics \textbar \ Harvard \& Smithsonian, 60 Garden Street, Cambridge, MA 02138, USA}
\email{cristilyn.watkins@cfa.harvard.edu}

\author[0000-0001-6588-9574]{Karen A. Collins}
\affiliation{Center for Astrophysics \textbar \ Harvard \& Smithsonian, 60 Garden Street, Cambridge, MA 02138, USA}
\email{karen.collins@cfa.harvard.edu}

\author[0000-0003-1361-985X]{Juliana Garc\'ia-Mej\'ia}
\affiliation{Kavli Institute for Astrophysics and Space Research, Massachusetts Institute of Technology, Cambridge, MA 02139, USA}
\affiliation{Center for Astrophysics \textbar \ Harvard \& Smithsonian, 60 Garden Street, Cambridge, MA 02138, USA}
\email{jgarciam@mit.edu}

\author[0000-0003-2171-5083]{Patrick Tamburo}
\affiliation{Center for Astrophysics \textbar \ Harvard \& Smithsonian, 60 Garden Street, Cambridge, MA 02138, USA}
\email{patrick.tamburo@cfa.harvard.edu}

\author[0000-0002-7733-4522]{Juliette Becker}
\affiliation{Department of Astronomy, University of Wisconsin–Madison, 475 N Charter St, Madison, WI 53706, USA}
\email{juliette.becker@wisc.edu}

\author[0000-0001-7254-4363]{Annelies Mortier}
\affiliation{School of Physics \& Astronomy, University of Birmingham, Edgbaston, Birmingham B15 2TT, UK}
\email{a.mortier@bham.ac.uk}

\author[0000-0001-8749-1962]{Thomas Wilson}
\affiliation{Department of Physics, University of Warwick, Gibbet Hill Road, Coventry CV4 7AL, UK}
\email{thomas.g.wilson@warwick.ac.uk}

\author[0000-0003-3623-7280]{Nicholas Scarsdale}
\affiliation{Department of Astronomy \& Astrophysics, University of California Santa Cruz, 1156 High Street, Santa Cruz, CA, 95064, USA}
\email{nscarsda@ucsc.edu}

\author[0000-0002-0388-8004]{Emily A. Gilbert}
\affiliation{Jet Propulsion Laboratory, California Institute of Technology, 4800 Oak Grove Drive, Pasadena, CA 91109, USA}
\email{emily.a.gilbert@jpl.nasa.gov}

\author[0000-0001-7047-8681]{Alex S. Polanski}
\altaffiliation{Percival Lowell Fellow}
\affil{Lowell Observatory, 1400 W Mars Hill Road, Flagstaff, AZ, 86001, USA}
\affil{Department of Physics and Astronomy, University of Kansas, Lawrence, KS 66045, USA}
\email{apolanski@lowell.edu}

\author[0000-0002-2532-2853]{Steve B. Howell}
\affiliation{NASA Ames Research Center, Moffett Field, CA 94035 USA}
\email{steve.b.howell@nasa.gov}

\author{Ian Crossfield}
\affiliation{Department of Physics and Astronomy, University of Kansas, Lawrence, KS, USA}
\email{ianc@ku.edu}

\author[0000-0001-6637-5401]{Allyson Bieryla}
\affiliation{Center for Astrophysics \textbar \ Harvard \& Smithsonian, 60 Garden Street, Cambridge, MA 02138, USA}
\email{abieryla@cfa.harvard.edu}

\author[0000-0002-5741-3047]{David R. Ciardi}
\affiliation{NASA Exoplanet Science Institute - Caltech/IPAC, 1200 E. California Blvd MS 100-22 Pasadena, CA 91125 USA}
\email{ciardi@ipac.caltech.edu}

\author[0000-0001-7139-2724]{Thomas Barclay}
\affiliation{NASA Goddard Space Flight Center, 8800 Greenbelt Road, Greenbelt, MD 20910, USA}
\email{thomas.barclay@nasa.gov}

\author[0000-0002-9003-484X]{David Charbonneau}
\affiliation{Center for Astrophysics \textbar \ Harvard \& Smithsonian, 60 Garden Street, Cambridge, MA 02138, USA}
\email{dcharbonneau@cfa.harvard.edu}

\author[0000-0001-9911-7388]{David W. Latham}
\affiliation{Center for Astrophysics \textbar \ Harvard \& Smithsonian, 60 Garden Street, Cambridge, MA 02138, USA}
\email{dlatham@cfa.harvard.edu}

\author[0000-0001-8898-8284]{Joseph M. Akana Murphy}
\altaffiliation{NSF Graduate Research Fellow}
\affiliation{Department of Astronomy \& Astrophysics, University of California Santa Cruz, 1156 High Street, Santa Cruz, CA, 95064, USA}
\email{jommurph@ucsc.edu}

\author[0000-0003-4150-841X]{Elisabeth Newton}
\affiliation{Department of Physics and Astronomy, Dartmouth College, Hanover, NH 03755, USA}
\email{elisabeth.r.newton@dartmouth.edu}

\author[0000-0001-8879-7138]{Bob Massey}
\affiliation{American Association of Variable Star Observers, 185 Alewife Brook Parkway, Suite 410, Cambridge, MA 02138, USA}
\email{bobmassey1@gmail.com}

\author[0000-0001-8227-1020]{Richard P. Schwarz}
\affiliation{Center for Astrophysics \textbar \ Harvard \& Smithsonian, 60 Garden Street, Cambridge, MA 02138, USA}
\email{rpschwarz@comcast.net}

\author[0000-0003-2163-1437]{Chris Stockdale}
\affiliation{Hazelwood Observatory, RMB 4036 Matta Drive, Hazelwood South, Hazelwood South, Victoria 3840, Australia}
\email{thestockdalefamily@bigpond.com}

\author[0000-0003-2127-8952]{Francis P. Wilkin}
\affiliation{Union College, 807 Union St., Schenectady, NY 12308}
\email{wilkinf@union.edu}

\author[0009-0009-0994-1767]{Roberto Zambelli}
\affiliation{Societa' Astronomica Lunae, Castelnuovo Magra Via Montefrancio 77 Italy}
\email{robertozambelli.rz@libero.it}

\shortauthors{Dattilo et al.}
\correspondingauthor{Anne Dattilo}
\email{adattilo@psu.edu}

\begin{abstract}
    Young planets with mass measurements are particularly valuable in studying atmospheric mass-loss processes, but these planets are rare and their masses difficult to measure due to stellar activity. We report the discovery of a planetary system around TOI-6109, a young, 75 Myr-old Sun-like star in the Alpha Persei cluster. It hosts at least two transiting Neptune-like planets 
    within 10-day orbital periods. Using three TESS sectors, 30 CHEOPS orbits, and photometric follow-up observations from the ground, we confirm the signals of the two planets. TOI-6109 b has an orbital period of P=$5.6904^{+0.0004}_{-0.0004}$ days and a radius of R=$4.87^{+0.16}_{-0.12}$ R$_\oplus$. The outer planet, TOI-6109 c has an orbital period of P=$8.5388^{+0.0006}_{-0.0005}$ days and a radius of R=$4.83^{+0.07}_{-0.06}$ R$_\oplus$.  These planets orbit just outside a 3:2 mean motion resonance. 
    The near-resonant configuration presents the opportunity to measure the planet's mass via TTV measurements and to bypass difficult RV measurements. Measuring the masses of the planets in this system will allow us to test theoretical models of atmospheric mass loss.
\end{abstract}

\section{Introduction}\label{sec:intro}
Demographic studies from Kepler show that small planets, with radii between 2--4\,\Re, are ubiquitous \citep{howard2012, petigura2018, hsu2019, kunimoto2020b, dattilo2023}, but studies of their masses show a striking range of compositions \citep{mayor2011, teske2021, luque2022, polanski2024}. Because we don't have any of these planets in our own Solar System, it is difficult to constrain their formation and evolution histories. We must observe planets in their infancy to understand their formation. Primordial versions of these planets are harder to detect for a variety of reasons, namely the relative rarity of their host stars and their high stellar activity.

Possible evolutionary histories of older, Kepler-like populations of planets can be traced via demographic features such as the radius valley. The radius valley is a region of low occurrence in the radius distribution that separates the larger sub-Neptune population from the smaller and rockier super-Earths \citep{fulton2017}. This feature can be replicated through a variety of physical processes, from formation causes to atmospheric mass loss to planet migration \citep[e.g.,][]{owen2017, rogers2021, gupta2020, venturini2020, lee2022, burn2024}. The timescales of these processes also vary, with most of them acting on relatively short timescales of less than 1\,Gyr. To catch a planet likely in the act of evolving, we must detect it at ages less than 100\,Myr.

In order to understand what processes dominate planet formation at early times, we need to catalog a large number of young planetary systems with periods, radii, and masses. One radius or mass value is degenerate with many compositions; e.g., see the ongoing discussion of the possibility of water worlds versus a single sub-Neptune population of H/He-dominated atmospheres \citep{luque2022, rogers2023, chakrabarty2024, parc2024}. A sample of planets with well-characterized masses and radii allows us to measure their composition at birth, while evolutionary processes are still ongoing. \added{The Transiting Exoplanet Survey Satellite \citep[\TESS; ][]{ricker2015}, presents the opportunity to study planets in well-aged stellar populations as it has observed $\sim$95\% of the sky.
The \TESS\ Hunt for Young and Maturing Exoplanets \citep[THYME; ][]{newton2019} collaboration, which focuses on detecting and characterizing exoplanets in young stellar groups and open clusters, has used \TESS\ to investigate early evolutionary processes.} Other work, such as \citet{fernandes2023} and \citet{vach2024} are using \TESS\ to do demographic studies of young planets as well.

In this paper we confirm another young planetary system, TOI-6109. The host is a young, 75\,Myr-old Sun-like star in the Alpha Persei cluster. With \TESS\ and \CHEOPS\ photometry we have confirmed two Neptune-sized planets orbiting at \perb\ and \perc\ days. This system is a near-laboratory experiment for understanding planet formation and evolution processes. These planets are in, or near, a near mean motion resonant orbital configuration, giving rise to observed large transit timing variations. Continued monitoring of this system will give us the information needed to measure their masses.

The paper is organized as follows. In Section~\ref{sec:data} we describe the data used for characterization of the star and planets. Section~\ref{sec:star} describes the host star properties. Section~\ref{sec:analysis} details our analysis and gives the planet parameters. Section~\ref{sec:discussion} discusses the planets as atmospheric targets. Finally, in Section~\ref{sec:conc} we give our summary and conclude.

\section{Data}\label{sec:data}
\subsection{Photometry}

\subsubsection{\TESS\ Photometry}

\TESS\ observed \thisstar\ in \added{years 1, 4, and 6} of its mission. In Sector 18 (UT 2019 November 02 to UT 2019 November 27) and Sector 58 (UT 2022 October 29 to UT 2022 November 26) it was observed with 2-minute cadence, \added{and in Sector 85 (UT October 26 to UT 2025 November 21) it was observed with 20-second cadence.} MIT's Quick Look Pipeline \citep{huang2020} detected a signal with a period of $8.5387\pm0.0001$ days and a radius of $4.05\pm0.31$\,\Re\ based on \added{the first two} sectors of data. The \TESS\ Science Office alerted it as a \TESS\ Object of Interest (TOI) on UT 2023 March 23, having passed all vetting tests. Subsequent to our planet search, a second transit signal was detected by QLP and alerted as a TOI on UT 2023 July 20 with a period of $5.6952\pm0.0005$ days and a radius of $3.92\pm0.46$\,\Re.

Meanwhile, after the alert of the first TOI, we performed our own search. We used a custom light curve extraction pipeline starting from the \TESS\ SPOC simple aperture photometry \citep[SAP;][]{twicken2010, morris2020}. We extracted the light curve and applied systematic corrections following the procedure in \cite{vanderburg2019}, and briefly summarize it here. The SPOC lightcurves were fit with a linear model, which consisted of the following components:

\begin{itemize}
    \item A basis spline (B-spline) with regularly spaced breaks at 0.2 day intervals to model long-term, low-frequency stellar variability
    \item The mean and standard deviation of the spacecraft quaternion time series within each light curve exposure
    \item Seven co-trending vectors from the SPOC Pre-search Data Conditioning's band 3 flux time series correction with the largest eigenvalues \citep{stumpe2012, stumpe2014, smith2012}
    \item A high-pass-filtered (0.1 days) flux time series for the SPOC background aperture
\end{itemize}

We searched the corrected light curves using \texttt{Notch \& LOCoR} \citep{Rizzuto2017}. We updated the Box-Least Squares (BLS) algorithm and search grid following \cite{Barber2024}. Using \texttt{Notch}, we detrended the light curve with a 0.5-day filtering window, removing stellar variability using a second-order polynomial while preserving trapezoidal, transit-like signals. We then searched for signals between 0.5 and 30 days using the BLS. We recovered four candidate signals; the 8-day previously alerted TOI, the 5-day second alerted TOI, and two other candidates near mean motion resonances at 11.5~days and 15.7~days that we can not yet confirm or disprove. 

\begin{figure*}
    \centering
    \includegraphics[width=\textwidth]{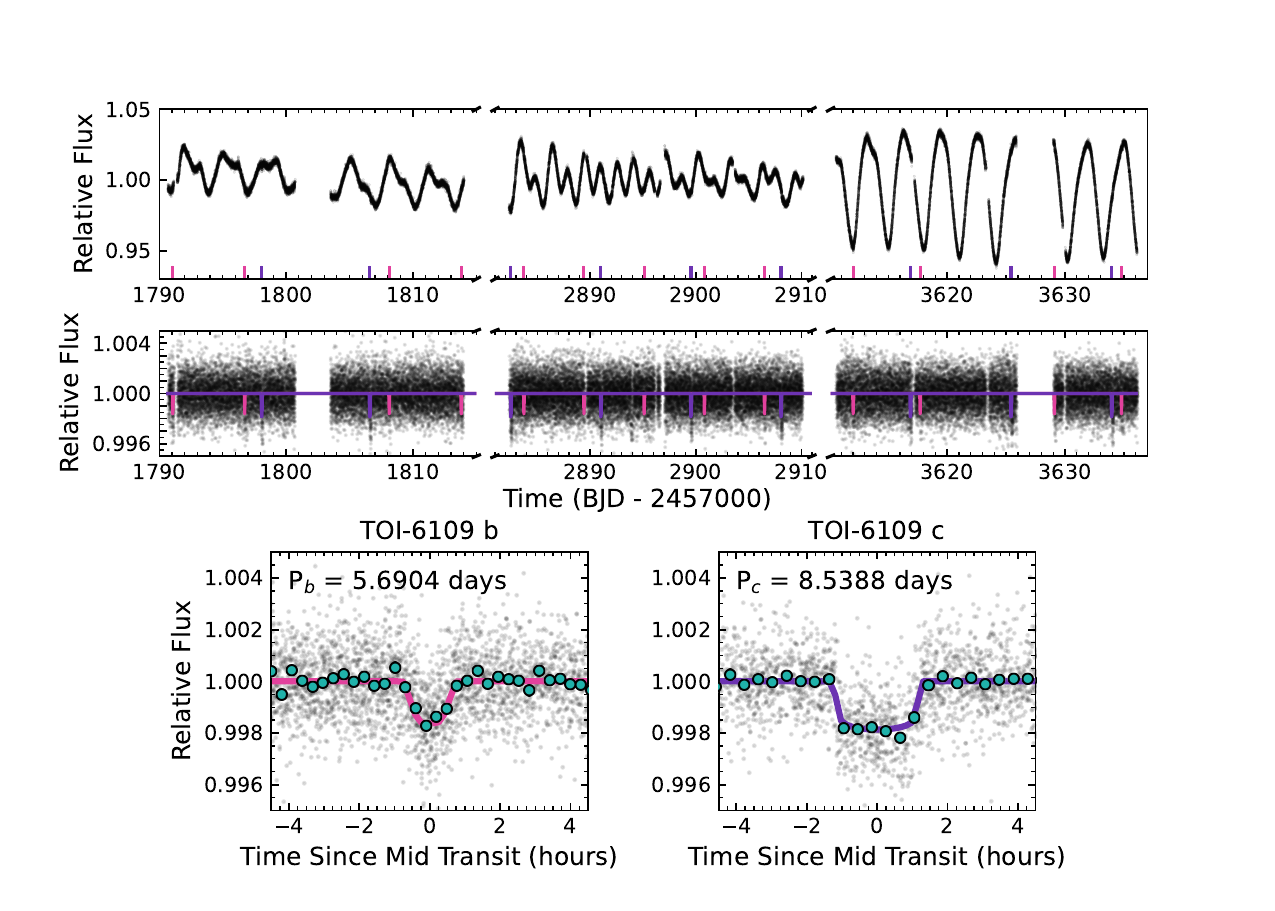}
    \caption{Top: unflattened \TESS\ lightcurve of TOI-6109. Middle: flattened lightcurve with two best-fit transit models for planets b and c. The colors of these lightcurves correspond to each of the planets in the bottom panels. Bottom: phase-folded lightcurves of planet b (left) and planet c (right). These include the additional \CHEOPS\ observations.}
    \label{fig:TESS-lc}
\end{figure*}

We also use our custom systematics-corrected light curves for the transit fitting analysis discussed in Section \ref{sec:transits}.
Prior our transit fitting process, outliers were removed from the light curves by fitting a Gaussian Process model and removing all data points more than 3-sigma from the mean. Light curve detrending was done simultaneously with transit fitting. The unflattened light curve, flattened lightcurve, and our transit fits can be seen in Figure~\ref{fig:TESS-lc}. 

\subsubsection{\CHEOPS\ Photometry}

\begin{deluxetable*}{ccccccc}
\tablecaption{CHEOPS observations\label{tab:cheops}}
\tablehead{
\colhead{ID} & \colhead{Start Date (UTC)} & \colhead{Dur (orbits)} & \colhead{File Key} & \colhead{Aperture (pixels)} & \colhead{RMS (ppm)} & \colhead{Planet}
}
\startdata
1 & 2023-10-06T04:26:58 & 8.13 & PR440018\_TG000101\_V0300 & 18 & 583 & c \\
2 & 2023-10-07T09:53:00 & 6.37 & PR440018\_TG000201\_V0300 & 18 & 495 & b \\
3 & 2023-10-18T18:56:58 & 7.01 & PR440018\_TG000202\_V0300 & 19 & 507 & b \\
4 & 2023-10-29T19:22:00 & 7.49 & PR440018\_TG000301\_V0300 & 16 & 595 & missed \\
5 & 2023-11-15T03:20:00 & 8.43 & PR440018\_TG000401\_V0300 & 18 & 502 & missed \\
6 & 2023-12-21T21:17:00 & 7.40 & PR440018\_TG000501\_V0300 & 19 & 564 & c \\
7 & 2023-12-25T21:49:52 & 8.69 & PR440018\_TG000601\_V0300 & 23 & 442 & missed$^a$ \\
8 & 2024-01-06T21:59:56 & 7.51 & PR440018\_TG000701\_V0300 & 18 & 476 & missed \\
\enddata
\tablenotetext{a}{These data overlapped with the \textit{Tierras} observation on UT 25-December-2023.}
\end{deluxetable*}

To refine the radii and ephemerides, we obtained 64 orbits of space-based observations with \CHEOPS\ DDT program CH\_PR440018 \added{(PI: A. Dattilo)}. This equaled two transits of planet~b, two transits of planet~c, and four non-detections. We summarize these observations in Table~\ref{tab:cheops}.

The first transit of \planetc\ was observed on UT 2023 October 06 with 8 \CHEOPS\ orbits. Because of data gaps due to the orbit of the telescope, the transit ingress was not observed, but egress was recovered. This transit was 3.65 hours early based on the period measured from \TESS\ data.
A second transit of \planetc\ was observed on UT 2023 December 21. The expected ephemeris of this transit was modified based on the first early-transit, but was still 15 minutes earlier than expected based on the best-fit period.

Two transits of \planetb\ were observed, with visits on UT 2023 October 07 and 2023 October 18. Both transits were earlier than expected (52.49 and 27.09 minutes early, respectively) based on the best-fit period from the initial \TESS\ ephemerides.
An additional visit on UT 2023 December 25 overlapped with an expected transit time of this planet; there is no evidence of ingress or egress in that data. Simultaneous observations were taken at Mt. Hopkins with the \textit{Tierras} telescope and we discuss those in Section~\ref{sec:ground}.

Additional visits were taken on UT 2023 October 29, 2023 November 15, and 2024 January 06 to search for the additional planet signals detected in our BLS search. All 3 were unsuccessful in observing planet transits. Our non-detection with \CHEOPS\ could either be because the planet candidates exhibit large TTVs and do not match the ephemerides measured from \TESS\ data (which would not be surprising given their near-resonant configuration) or because the planet candidates are false alarms. More data will be required to confirm or refute these candidates and we defer that analysis to future work. 

The data were processed through the \CHEOPS\ Data Reduction Pipeline \citep[DRP;][]{hoyer2020}. We chose the aperture with the smallest 10 min-CDPP in the DRP photometry to use for each observation's light curve.
Before transit fitting, each light curve was detrended with a linear least-squares model to remove trends from the orbital motion of the spacecraft. This model included sine and cosine components for the roll angle of the spacecraft during each visit. The stellar activity signal was left in the light curve to be detrended simultaneously with the transit model.

\subsubsection{Ground-based photometric follow-up}\label{sec:ground}
\paragraph{LCOGT}

We observed two transit windows of \planetc\ in Pan-STARRS $z_s$ band and eight transit windows of \planetc\ in Sloan $i'$ band from the Las Cumbres Observatory Global Telescope (LCOGT) \citep{brown2013} 1\,m network nodes at McDonald Observatory near Fort Davis, Texas, United States (McD), and Teide Observatory on the island of Tenerife (TEID). See Table~\ref{tab:obs_lco_midtime} for details of each observation. The 1\,m telescopes are equipped with a $4096\times4096$ SINISTRO camera having an image scale of $0\farcs389$ per pixel, resulting in a $26\arcmin\times26\arcmin$ field of view. The images were calibrated by the standard LCOGT {\tt BANZAI} pipeline \citep{mccully2018} and differential photometric data were extracted using {\tt AstroImageJ} \citep{collins2017}. We used circular photometric apertures that excluded all of the flux from the nearest known neighbor in the Gaia DR3 catalog (Gaia DR3~241035802330976896), which is $5\farcs8$ east of \thisstar. Due to the large transit timing variations in the system, our LCOGT observing windows were not long enough to ensure that we captured an actual full transit at each observed epoch. Nevertheless, we extracted the timing of any apparent ingress and/or egress signals that appeared to be consistent with the expected transit depth, and for full transits, also consistent with the expected duration. The transit center times corresponding to each apparent detected event are listed in Table~\ref{tab:obs_lco_midtime}. \added{As our observational baseline has increased, the best-fit average period has changed and retroactively moved the expected ephemerides.} Because some of the apparent transit times appear to be inconsistent with the TTV models implied by our higher precision space-based data, we elected to exclude the LCOGT transits from our analysis in Section~\ref{sec:analysis}. 

\paragraph{Tierras} 
We observed \thisstar\ from the ground with the \textit{Tierras} Observatory, a 1.3-m ultra-precise fully-automated photometer located within the Fred Lawrence Whipple Observatory (FLWO) atop Mt Hopkins, Arizona \citep{garciamejia2020}. We gathered light curves with \textit{Tierras} through its custom filter ($\lambda_C = 863.5$\,nm, $\Delta \lambda = 40$\,nm at FWHM) during the nights of UT 2023 November 2, November 5, December 15, December 22, December 26, and 2024 January 30. We collected 40-second exposures of the target across the six dates, obtaining 106, 674, 306, 541, 481, and 364 individual images, respectively. The airmass probed by the observations across all the dates ranged from 1.02 to 1.67. The observing conditions were as follows: November 2 and November 5 were photometric, with clear conditions and a median seeing of 1.5'' and 1.65''; December 15 was mostly clear, with some passing clouds and a median seeing of 2.3''; December 22 was also photometric, with clear conditions for the entire observing run and a median seeing of 1.9''; December 26 was clear up until the end of the evening, when significant clouds affected the observations; 2024 January 30 was affected by passing clouds throughout the entire run. The last two nights have median seeing values of 2.11'' and 1.55'', respectively.     

The \textit{Tierras} data were reduced, and the photometry extracted, via the \textit{Tierras} custom data reduction pipeline. In broad terms, the pipeline bias-subtracts and stitches the images, performs aperture photometry using the \texttt{photutils} library \citep{bradley2024}, and employs an interactive weighting scheme to automatically select and weight reference stars in the field based on the measured scatter in their corrected fluxes (see \citet{tamburo2022} for details).

The 6.31 hour observation on December 26 contained a significant asymmetric dip in the light curve. This observation coincided with a \CHEOPS\ visit that began on UT 2023 December 25 and was near a predicted transit time for \planetb. The feature in the lightcurve is deeper than transits for \planetb\ measured by \TESS\ or \CHEOPS, and remains asymmetric after reduction. Unfortunately, it falls within a gap in the \CHEOPS\ data where there is little evidence of ingress or egress. \added{If the feature is a full transit of \planetb, it would be 61.1~minutes early.}

\subsection{High-resolution imaging}
Close stellar companions (bound or line of sight) can confound exoplanet transit discoveries in a number of ways \citep[e.g.,][]{ciardi2015, furlan2017}.  The detected signal might be a false positive due to a background eclipsing binary and even real planet discoveries will yield incorrect stellar and exoplanet parameters if a close companion exists and is unaccounted for. \citet{lester2021} have also shown that the presence of a close companion star leads to the non-detection of small planets residing with the same exoplanetary system. Since about one-half of all FGK stars are binary or multiple \citep[e.g.,][]{matson2017} high-resolution imaging provides a deep look into the stellar system and can detect and yield information on close (sub)stellar companions. We use both adaptive optics and speckle imaging to search for companions.

\subsubsection{Adaptive Optics}
To rule out possible contamination by closely bound or line-of-sight companions, we observed TOI-6109 with NIRC2 adaptive optics (AO) imaging at Keck Observatory. Observations were made on UT 2023 August 5 behind the natural guide star systems \citep{dekany2013, wizinowich2000} in the narrow band J continuum (\texttt{Jcont}; $\lambda_o = 1.2132; \Delta\lambda = 0.0198~\mu$m) and the narrow band K continuum (\texttt{Kcont}; $\lambda_o = 2.2706; \Delta\lambda = 0.0296~\mu$m) filters. Narrow angle mode was used, providing a pixel scale of around 0.01\arcsec~px$^{-1}$ and full field of view of about 10\arcsec. A standard three-point dither pattern was used to avoid the lower-left quadrant of the detector, which is typically noisier than the other three quadrants. The dither pattern has a step size of 3\arcsec. Each dither position was observed three times, with 0.5\arcsec~positional offsets between each observation, for a total of nine frames. The reduced science frames were combined into a single mosaiced image with a final resolutions of 0.04\arcsec~ and 0.05\arcsec~and at \texttt{Jcont} and \texttt{Kcont}, respectively.

The sensitivity of the final combined AO image were determined by injecting simulated sources azimuthally around the primary target every $20^\circ $ at separations of integer multiples of the central source's FWHM \citep{furlan2017}. The brightness of each injected source was scaled until standard aperture photometry detected it with $5\sigma $ significance.  The final $5\sigma $ limit at each separation was determined from the average of all of the determined limits at that separation and the uncertainty on the limit was set by the root-mean-squared dispersion of the azimuthal slices at a given radial distance.
No other nearby stellar companions are identified within our detection limits.

\subsubsection{Speckle Imaging}

TOI-6109 was observed on 2022 December 02 UT using the optical high-resolution speckle instrument ‘Alopeke on the Gemini North 8-m telescope \citep{scott2021}. ‘Alopeke provides simultaneous speckle imaging in two bands (562~nm and 832~nm) with output data products including a reconstructed image with robust magnitude contrast limits on companion detections. Five sets of 1000$\times$0.06 second images were obtained and processed with our standard reduction pipeline (see \citet{howell2011}).
We find that TOI-6109 is a single star with no close companion detected brighter than 5-8.5 magnitudes below that of the target star and within the 8-m telescope diffraction limit (20~mas) out to 1.2”. At the distance of TOI-6109 (d=148\,pc) these angular limits correspond to spatial limits of 3 to 182~au.

\subsection{Spectroscopy}
\paragraph{TRES}
We observed TIC 384984325 using the Tillinghast Reflector Echelle Spectograph (TRES) mounted on the 1.5 meter telescope at FLWO on Mt. Hopkins, AZ. The data were collected with a resolving power of $\lambda / \Delta\lambda = 44,000$ and moderate signal-to-noise, ranging from 25 to 35 per resolution element. 
We obtained three observations of TIC 384984325. The observations showed no evidence for large radial velocity variations, ruling out eclipsing binary stars as a possible explanation for the transit signal we see. 
The velocities from the three TRES observations yield an RMS scatter of 129 m/s, but we expect a large amount of scatter due to the stellar activity from the young star.
Spectroscopic parameters for the two stars were derived using the Stellar Parameter Classification (SPC) code \citep{buchhave2012, buchhave2014}, with the results presented in Table~\ref{tab:prop}.

\paragraph{McDonald Observatory}

We observed TIC 384984325 with the 2.7\,m Harlan J. Smith Telescope (HJST) at the McDonald Observatory, using the TS23 configuration of the Robert G. Tull Coud\'e spectrograph \citep{Tull95}. This instrument provides high-resolution spectroscopy with $R=60000$ in a non-contiguous spectroscopic range from 3400 and 10900~\AA, including the H$\alpha$ and Li 6708 \AA~lines, which are useful for establishing stellar youth. This setup provided an SNR that peaked at $\sim$50 per resolution element, which was sufficient to measure line strengths and a radial velocity. 

This spectrum was reduced using a publicly available reduction pipeline designed for the Tull Coud\'e Spectrograph\footnote{\url{https://tull-coude-reduction.readthedocs.io/en/latest/}}. We then used the spectral line broadening functions from the \texttt{saphires} package to compute a radial velocity \citep{tofflemire2019}, and also fit the Li 6708 \AA~and H$\alpha$ lines with Gaussian profiles. No H$\alpha$ emission was detected, but a strong Li line was present, which is an indicator of stellar youth. We include these observations in our overview of the host star in Table \ref{tab:prop}. 
 
\section{Host Star Properties}\label{sec:star}

\thisstar\ is a young, nearby G star in the Alpha Persei open cluster. We present the stellar parameters measured in Gaia DR3 \citep{gaiadr3}, as well as our best-fit values from ground-based observing in Table~\ref{tab:prop}.
We use these stellar properties to calculate the planet properties throughout the rest of the work.

\begin{deluxetable*}{lccc}[htb!]
\label{tab:prop}
\centering
\tabletypesize{\scriptsize}
\tablewidth{0pt}
\tablecaption{Properties of the host star \thisstar.}
\tablehead{\colhead{Parameter} & \colhead{Value} & \colhead{Source} }
\startdata
    TIC ID & 384984325 & TESS Input Catalog\\
    TOI ID & 6109 & \citet{guerrero2021}\\
    \textit{Gaia} DR3 ID & 241035596174886016 & Gaia DR3\\
    TYC & 2869-2813-1 & \citet{hoogerwerf2000}\\
    \hline
    \multicolumn{3}{c}{Astrometry}\\
    \hline
    $\alpha$ & 03:20:31.6229965872 & Gaia DR3  \\
    $\delta$ & +42:36:25.486570296 & Gaia DR3  \\
    $\mu_\alpha$ (mas yr$^{-1}$) & $27.015\pm0.020$ & Gaia DR3  \\
    $\mu_\delta$ (mas yr$^{-1}$) & $-28.612\pm0.019$ & Gaia DR3 \\
    $\pi$ (mas) & $6.7335\pm0.0182$ & Gaia DR3 \\
    distance (pc) & $148.5\pm0.4$& Gaia DR3 \\
    \hline
    \multicolumn{3}{c}{Photometry}\\
    \hline
    Spectral Type & G3 & \citet{zhang2023}\\
    G$_{Gaia}$ (mag) & $10.819428\pm0.003615$ & Gaia DR3 \\
    BP$_{Gaia}$ (mag) & \added{$11.1789\pm{0.0082}$} & Gaia DR3 \\
    RP$_{Gaia}$ (mag) & \added{$10.2908\pm{0.0060}$} & Gaia DR3 \\
    B$_T$ (mag) & $11.64\pm0.08$ & Tycho-2 \\
    V$_T$ (mag) & $11.05\pm0.08$ & Tycho-2 \\
    J (mag) & $9.656\pm0.023$ & 2MASS \\
    H (mag) & $9.316\pm0.031$  & 2MASS \\  
    Ks (mag) & $9.251\pm0.022$  & 2MASS \\
    \hline
    \multicolumn{3}{c}{Kinematics \& Galactic Position}\\
    \hline
    RV$_{\rm{Bary}}$  (km\, s$^{-1}$) & $2.3329\pm0.1105$& Gaia DR3 \\
    U (km\, s$^{-1}$) & $-14.27\pm0.09$ & This Work\\ 
    V (km\, s$^{-1}$) & $-22.91\pm0.05$ & This Work\\ 
    W (km\, s$^{-1}$) & $-6.66\pm0.03$ & This Work\\ 
    X (pc) & $-125.77\pm0.34$ & This Work \\ 
    Y (pc) & $72.40\pm0.20$ & This Work \\ 
    Z (pc) & $-31.57\pm0.09$ & This Work \\ 
    \hline
    \multicolumn{3}{c}{Physical Properties}\\
    \hline
    $P_{\rm{rot}}$ (days) & $3.02^{+0.02}_{-0.02}$ & This Work\\
    $v \sin i$ \ (km\,s$^{-1}$) & $18.2\pm0.5$ & This work\\ 
    $f_{bol}$\,(erg\,cm$^{-2}$\,s$^{-1}$)& ($0.128\pm0.009)\times10^{-8}$ & This Work\\ 
    Age (Myr) & $75\pm5$ & \citet{galindogull2022}\\
    T$_{\mathrm{eff}}$ (K) & $5660\pm50$ & This work\\ 
    M$_\star$ (M$_\odot$) & $1.03\pm0.05$ & This Work \\
    R$_\star$ (\Rsun{}) & $1.021\pm0.038$ & This Work\\ 
    L$_\star$ (\Lsun{}) & $0.882\pm0.055$ & This Work \\
    $\rho_\star$ ($\rho_\odot$) & $1.369\pm0.484$ & This Work\\ 
    $\log{(g)}$ ($\log(\mathrm{cm/s^2})$) & $4.45\pm0.10$ & This work \\
    EW(Li) (m\r{A})& $193\pm6$ & This Work \\
\enddata
\end{deluxetable*}

\begin{deluxetable}{lccccc}
\tablecaption{Apparent LCOGT \& Tierras Transit Times}
\label{tab:obs_lco_midtime}
\tablehead{\colhead{Obs. Date} & \colhead{T$_{\rm c}$} & \colhead{T$_{\rm c}$ O-C$^b$} & \colhead{Apparent} & \colhead{Instrument}\\   
\colhead{UTC} & \colhead{BTJD$^a$} & \colhead{} & Event & \colhead{}}
\startdata
\textbf{TOI-6109\,b }  &  & & & \\
$\:$ $\:$   2023-09-09 & 3196.7674 & +2.73 &	 egress   & LCO-McD  \\
$\:$ $\:$   2023-09-26 & 3213.8225 & +2.34 &	 full   & LCO-McD  \\
$\:$ $\:$   2023-10-01 & 3219.5287 & +2.72 &	 full   & LCO-TEID  \\
$\:$ $\:$   2023-11-05 & 3253.6715 & +2.72 &	 full   & LCO-TEID  \\ 
$\:$ $\:$   2023-11-22 & 3270.7689 & +3.34 &	 egress   & LCO-McD  \\
$\:$ $\:$   2023-11-27 & 3276.4405 & +2.89 &	 full   & LCO-TEID  \\
$\:$ $\:$   2023-12-14 & - & &	 no event  & LCO-TEID  \\
$\:$ $\:$   2023-12-25 & 3304.7275 & -1.01 &   full 	&  Tierras  \\
$\:$ $\:$   2024-08-21 & - & &	 no event  & LCO-McD  \\
\textbf{TOI-6109\,c }  & & & & \\
$\:$ $\:$   2023-09-02 & 3189.7975 & -2.16 &	 egress   & LCO-McD$^c$  \\
$\:$ $\:$   2023-11-17 & 3266.5987 & -3.34 &	 full     & LCO-TEID$^d$  \\
\enddata
\tablenotetext{a} {BJD$_{\rm UTC}-2457000$}
\tablenotetext{b} {time early (-), late (+), in hours, relative to linear ephemerides in Table \ref{tab:transfit}}
\tablenotetext{c} {McDonald Observatory near Fort Davis, Texas, United States (McD)}
\tablenotetext{d} {Teide Observatory on the island of Tenerife (TEID)}
\end{deluxetable}

%

\subsection{Fundamental Stellar Parameters}
To derive stellar properties, we start by fitting the spectral energy distribution (SED) of TOI-6109 following the approach outlined in \citet{mann2015, mann2020}. 
We combine photometry from 2MASS and WISE and stellar template with Phoenix BT-SETTL models \citep{allard2012} to cover wavelength gaps, resulting in an absolutely-calibrated spectrum. From this spectrum, we compute $F_{bol}$ by integrating the spectrum with wavelength and derive stellar luminosity ($L_*$) using the \textit{Gaia} DR3 parallax. The effective temperature ($T_{eff}$) is estimated by comparing the calibrated spectrum to atmospheric models. We then determine stellar radius ($R_*$) using the Stefan-Boltzmann relation.
To estimate stellar mass ($M_*$), we fit evolutionary models from the Dartmouth Stellar Evolution Database \citep[DSEP; ][]{dotter2008} and the PARSEC stellar evolution tracks \citep{bressan2012} taking the average between fits.

The final error analysis accounts for errors in template choice, systematics in estimating $T_{eff}$, shape errors in the templates, as well as uncertainties in parallax and observed photometry. The uncertainty on stellar mass comes from the difference between the DSEP and PARSEC fits. The final fit yielded $f_{bol}=(0.128\pm0.009)\times10^{-8}$\,(erg\,cm$^{-2}$\,s$^{-1}$), $L_*=0.882\pm0.055$\,\Lsun, $T_{eff}=5660\pm50$\,K, $R_*=1.021\pm0.038$\,\Rsun, and $M_*=1.03\pm0.05$\,M$_{\odot}$. These parameters are consistent with those in the \TESS\ Input Catalog.

\subsection{Association to Alpha Per}
\thisstar\ was first identified as a candidate member of the Alpha Per cluster in \citet{hoogerwerf2000}, who used proper motions to identify additional candidates from the Tycho Reference Catalog \citep{hog1998}. It has since then remained a strong member candidate as new data have become available from Gaia.
We substantiate its membership by comparing its galactic position, kinematics, Lithium abundance, and rotation with other cluster members.

\subsubsection{Galactic Position}
\begin{figure*}
    \centering
    \includegraphics[width=\textwidth]{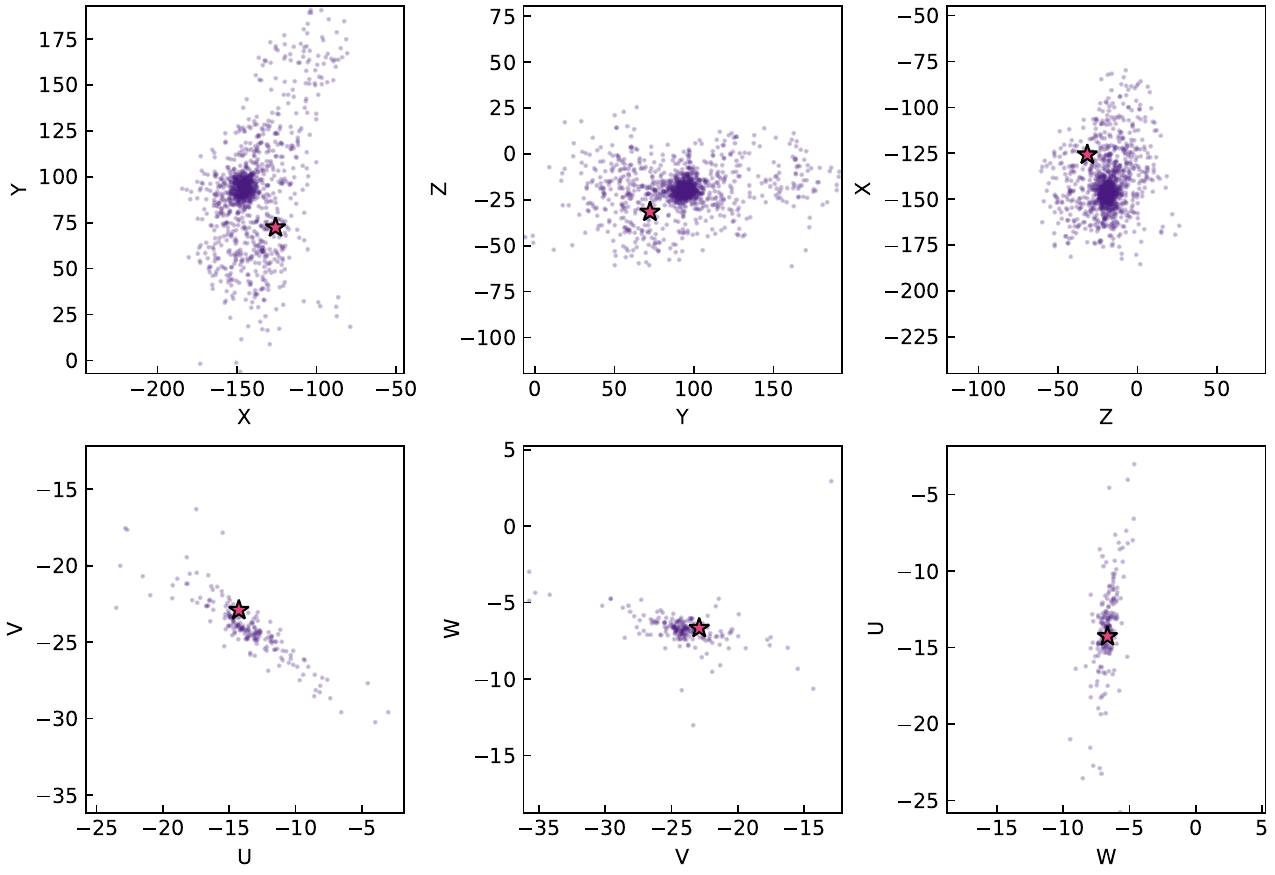}
    \caption{Relative position and velocity of \thisstar\ (pink star) relative to cluster members identified by \citet{meingast2021} (purple dots). The top row shows the position (X, Y, Z) and the bottom row shows the galactic space velocity (U, V, W), both calculated from Gaia DR3 measurements. This star is in the corona of the cluster but is in good agreement with the average kinematic velocity of the cluster.}
    \label{fig:membership}
\end{figure*}

\citet{meingast2021} identified this star as a member of the stellar corona of Alpha Per using Gaia DR2 data and accounting for cluster bulk velocities and spatial distribution. We use Gaia DR3 kinematic and positional measurements to calculate Cartesian Galactic coordinates as (X, Y, Z) = $(-124.99\pm0.34, 71.96\pm0.20, -31.38\pm0.09)$ pc and cylindrical Galactocentric coordinates as (R, $\phi$, z) = $(8226.14\pm0.34\,\textrm{pc}, 179.96\pm0.01\degree, -6.18\pm0.8\,\textrm{pc})$.
While this star is not in the cluster center, it is in the corona (Figure~\ref{fig:membership}, top row).

\subsubsection{Kinematics}
The Galactic space velocity values were calculated using the Python package \texttt{PyAstronomy}\footnote{\url{https://pyastronomy.readthedocs.io/en/latest/pyaslDoc/aslDoc/gal_uvw.html}}, which is based on the methodology from \citet{johnson1987}, without correction for solar motion. Based on proper motion and parallax measurements from Gaia DR3, the Galactic velocity of the star is (U, V, W) = $(-14.27\pm0.09, -22.91\pm0.05, -6.66\pm0.03)$ km\,s$^{-1}$. This agrees with the mean velocity of stars in the cluster measured by \citet{meingast2021} from Gaia DR3 values (Figure~\ref{fig:membership}, bottom row).

\subsubsection{Rotation}
We measure the rotation period of the star to be $3.02\pm0.02$ days from our Gaussian Process fitting of the lightcurve (Section~\ref{sec:analysis}). This is in good agreement with the rotation-temperature sequence for the cluster \citep[see Figure 4][]{boyle2023}. We determine the gyrochronal age using \texttt{gyro-interp} \citep{bouma2023}, which interpolates between open cluster rotation sequences. This yields an age of $56.72^{+46.13}_{-38.72}$ years. Since the interpolation has a lower limit of 80\,Myr, this result corresponds to the polynomial rotation period value set by the Alpha Per cluster data and effectively serves as an upper limit.

\subsubsection{Lithium Sequence}
\begin{figure*}
    \centering
    \includegraphics[width=\textwidth]{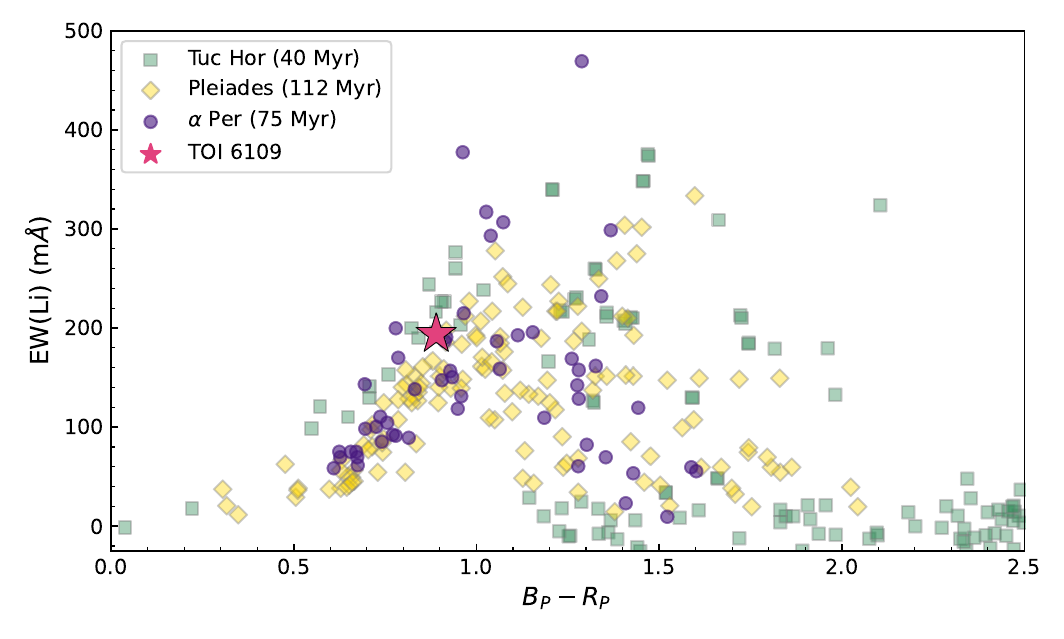}
    \caption{Lithium equivalent widths of Alpha Per cluster members (purple circle) from \citet{balachandran2011} and \citet{galindoguil2022} plotted as a function of Gaia $B_P - R_P$ color. \thisstar\ (pink star) agrees with other Alpha Per cluster members. Two other well-studied open cluster Li sequences are plotted; Tucana-Horologium \citep[40\,Myr, green square, data from][]{kraus2014} and the Pleiades \citep[112\,Myr, yellow diamond, data from][]{bouvier2018}. Because all three clusters overlap in the parameter space of \thisstar\ its age must lie between the two.}
    \label{fig:lithium}
\end{figure*}

We empirically confirm the cluster age and this star's membership by comparing the Li sequence to benchmark associations. Lithium equivalent widths as a function of color can be compared to associations with known ages to determine an age \citep{soderblom2014}. 
To create the lithium sequence of Alpha Per we take the stellar samples with lithium equivalent widths from \citet{balachandran2011} and \citet{galindoguil2022} and cross-match with Gaia to obtain their $B_P - R_P$ color. This lithium sequence is plotted in purple in Figure~\ref{fig:lithium}. 
We measure the lithium equivalent width (from the Li\,I\,6708\r{A} line) of \thisstar\ to be $193\pm6$\,m\r{A}. \added{We did not deblend the EW with the weak Fe I line at 6708.4\,\r{A}. The comparison stars from Alpha Per also do not take the blend into account for their measurements}. Our equivalent width is in agreement with literature values at similar colors for this cluster.

Empirically, we determine the age of the cluster by comparing the Li sequence to two other associations. Figure~\ref{fig:lithium} shows the Tucana-Horologium association \citep[40\,Myr, green square, data from][]{kraus2014} and the Pleiades \citep[112\,Myr, yellow diamond, data from][]{bouvier2018}. Alpha Per lies between these associations and the Li sequence agrees with the $75\pm5$\,Myr age from \citet{galindoguil2022}. \added{\citet{galindoguil2022} found their age by locating the lithium depletion boundary for the cluster and then deriving an evolutionary age based on that from several different theoretical models.}
Using \texttt{EAGLES} \citep{jefferies2023}, which models the age probability distribution based on a sample of stars in open clusters, we estimate the star's age from its equivalent width and effective temperature to be $71.21^{+56.9}_{-44.1}$\,Myr.

\added{\subsubsection{Galactic Traceback}
To see if this star is consistent with a birth location with other cluster members, we conducted a galactic traceback analysis using the methodology of \citet{kerr2022}. We used the package \texttt{galpy} \citep{bovy2015}, which uses asteroseismic and kinematic inputs, including RA, Dec, distance, proper motions, and RVs. We compared the trajectory of this star to the other cluster members used previously in this section. Our results indicate that the inferred birth position of this star is consistent with that of the core cluster members identified by \citet{meingast2021}.}

\subsection{Inclination of Stellar Rotation Axis}
Using our knowledge of the star, including its radius, rotation period, and $v \sin i$, it is possible to test whether the stellar spin and planetary orbit are aligned. To calculate the inclination of the star we use the formalism of \citet{masuda2020}. The calculated stellar inclination is consistent with alignment with the planet, yielding a limit of $>81.9\degree$ at 68.5\% confidence and $>76.1\degree$ at 95\% confidence, consistent with aligned orbits for the planets.

\section{Analysis}\label{sec:analysis}

\subsection{Planetary parameters (transit fitting)}\label{sec:transits}
To derive planet parameters, we use only space-based data: our custom systematics-corrected lightcurves for \added{all three} \TESS\ sectors and eight \CHEOPS\ lightcurves with the spacecraft's orbital motion removed. All data undergo an outlier removal process. Due to the star's youth, there is substantial ($>40$ppt) stellar variability.
We account for this variability using a Gaussian Process (GP) model implemented via \texttt{celerite} \citep{dfm2014}, simultaneously fitting both planet transit models \citep[via \texttt{batman};][]{kreidberg2015}.
The GP employs a Simple Harmonic Oscillator kernel with the parameters $S, Q, \omega_{GP}$. \added{We fit two linear limb-darkening coefficients for each space telescope, $u_{\textrm{TESS}, 1}, u_{\textrm{TESS}, 2}, u_{\textrm{CHEOPS}, 1}; u_{\textrm{CHEOPS}, 2}$, and the stellar density $\rho_*$. For each planet, we jointly fit the parameters the planet-to-star-radius ratio, $p$, and impact parameter, $b$, following \citet{espinoza2018}. Because the system exhibits TTVs, we cannot use a linear ephemeris to model the planet transits. Instead, we fit each transit time individually, $t_N$, for a total of 37 parameters.} Orbital period is derived from a linear fit to each planet's transit times.
For our initial fit, we assume a circular orbit ($e=0$) and report the resulting parameters in the lefthand columns of Table~\ref{tab:transfit}..

\begin{figure*}
    \centering
    \includegraphics[width=\textwidth]{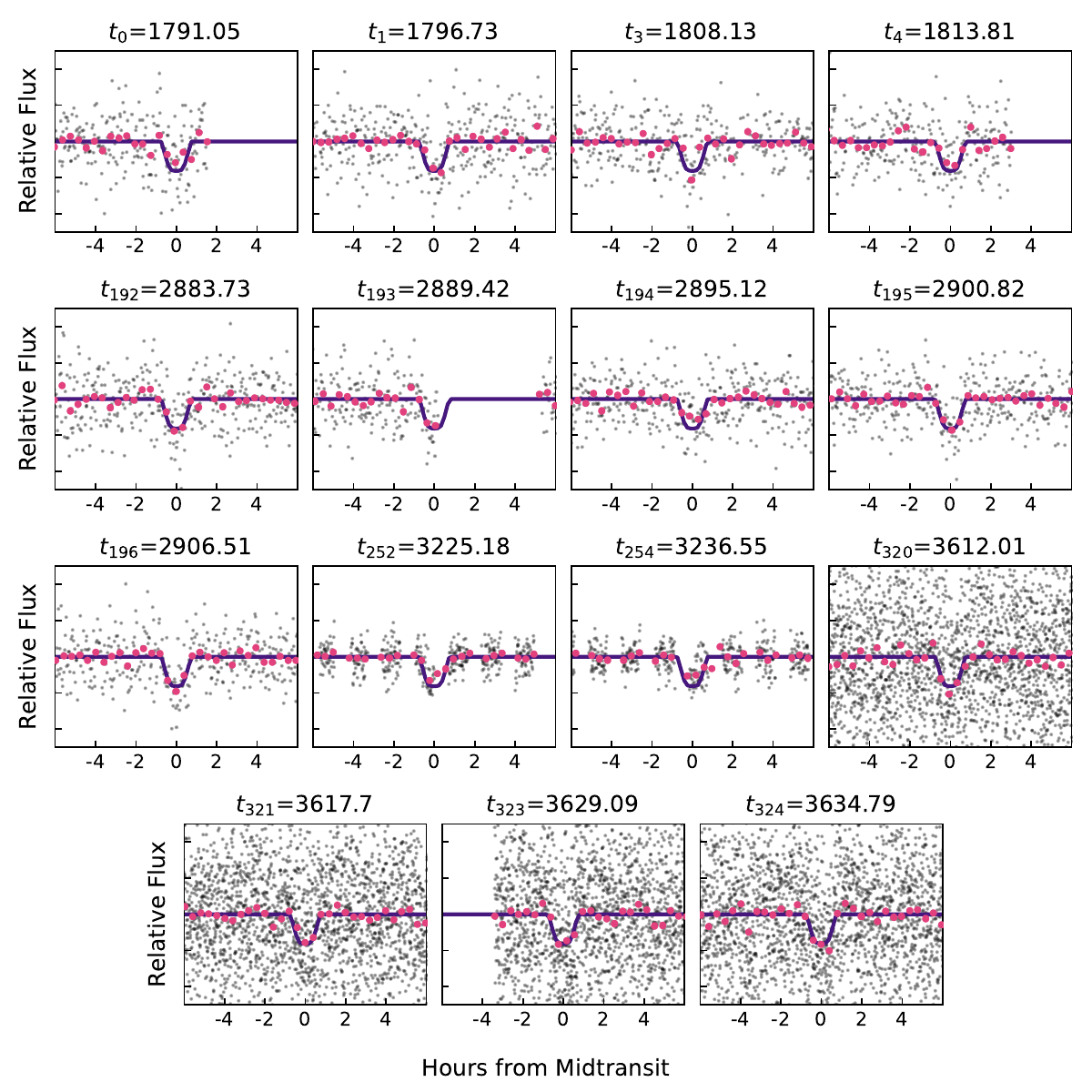}
    \caption{\added{Individual transits for planet~b. The transit model is in black and 30-minute binned points are shown in pink. The first four transits (top row) are from \TESS\ Sector 18 while the next five are from Sector 58. The middle two transits in the third tow are from two \CHEOPS\ visits, followed by four transits from Sector 85 with 20-second cadence observations.}}
    \label{fig:5day-transits}
\end{figure*}

\begin{figure*}
    \centering
    \includegraphics[width=\textwidth]{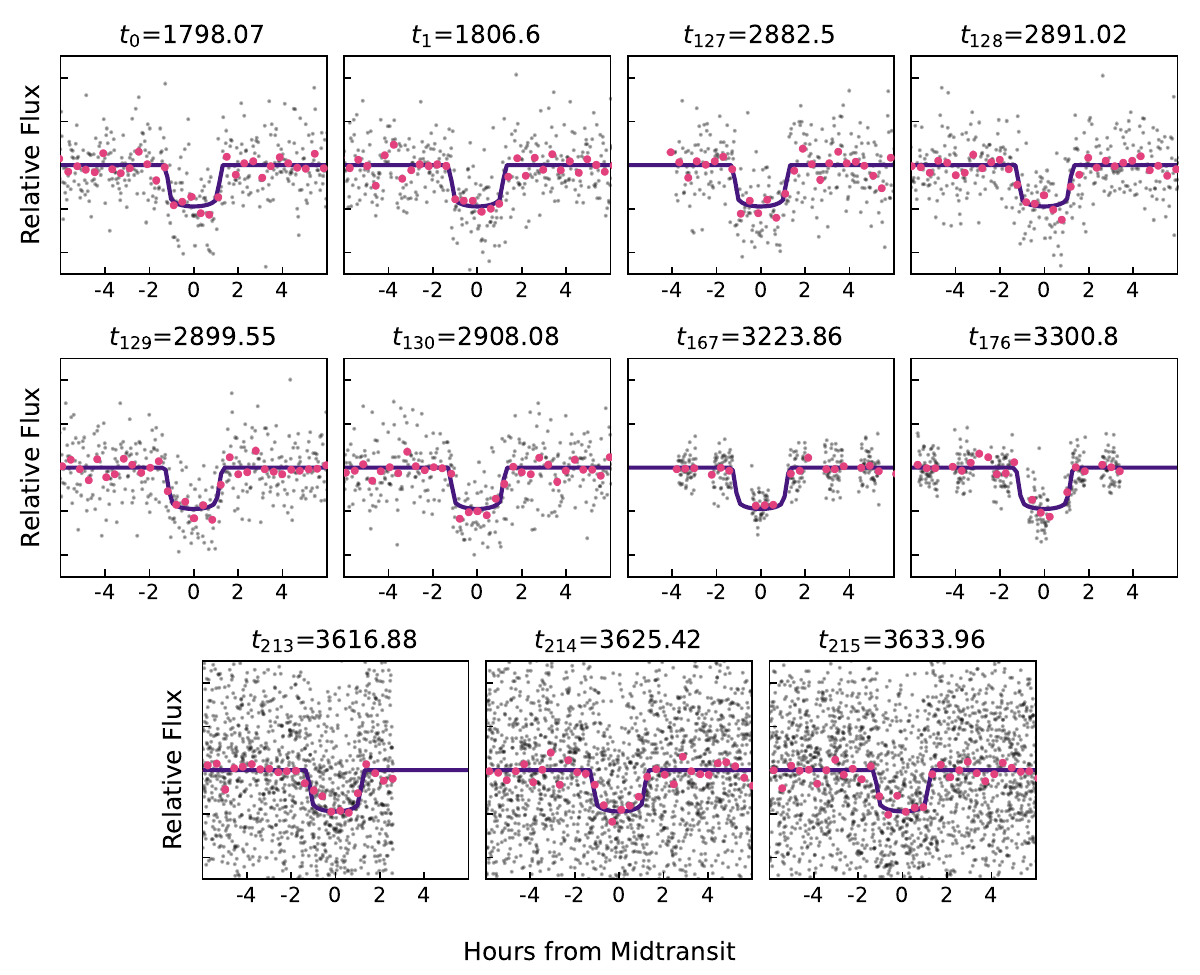}
    \caption{\added{Individual transits for planet~c. The transit model is in black and 30-minute binned points are shown in pink. The first two transits are from \TESS\ Sector 18, the next four from Sector 58, and the following two transits from two \CHEOPS\ visits, with the last row showing three transits from Sector 85.}}
    \label{fig:8day-transits}
\end{figure*}

To model the system, we include all transits in our light curve model and normalize the model flux to zero. We then subtract this model from the observed light curve flux, leaving only the residuals. Finally, we fit the GP to these residuals. For parameter estimation, we use \texttt{emcee} \citep{dfm2014} as our MCMC sampler, treating the GP as our likelihood function. \added{To ensure convergence, we ran the model with 100 walker until it reached 50$\times$ the autocorrelation time, $\tau$, then ran the sampler for an additional 10$\times \tau$ (53,000 steps). We also check convergence with the Gelman-Rubin statistic, confirming that all 37 parameters satisfy the convergence criterion R $<$ 1.05.}
The priors implemented for both planets are listed in Table~\ref{tab:priors}. The results of our fit can be seen in the flattened lightcurve and transit models shown in the middle row of Figure~\ref{fig:TESS-lc}, as well as the phase-folded light curves for both planets in the bottom row. Each individual transit for our \TESS\ and \CHEOPS\ observations are shown in Figures~\ref{fig:5day-transits} and \ref{fig:8day-transits}.

\begin{deluxetable}{cc}\label{tab:priors}
\tablecaption{Transit Fit Priors}
\tablehead{
    \colhead{Parameter} & \colhead{Prior}
}
\startdata
$T_0$ (TJD) & $\mathcal{U}[T_i - 0.2,T_i + 0.2]$ \\
$p$ & $\mathcal{U}[0, 1]$ \\
$b$ & $\mathcal{U}[0, 1]$ \\
$\rho_{\star}$ ($\rho_{\odot}$) & $\mathcal{N}[1.41, 0.20]$ \\
$u_{\textrm{TESS}, 1}$\,$^a$ & $\mathcal{N}[0.1938, 0.20]$ \\
$u_{\textrm{TESS}, 2}$\,$^a$ & $\mathcal{N}[0.3273, 0.20]$ \\
$u_{\textrm{CHEOPS}, 1}$\,$^b$ & $\mathcal{N}[0.5, 0.10]$ \\
$u_{\textrm{CHEOPS}, 2}$\,$^b$ & $\mathcal{N}[0.15, 0.10]$ \\
$S$ & $\mathcal{U}[-50,20]$ \\
$Q$ & $\mathcal{U}[-50,20]$ \\
$\omega_{GP}$ & $\mathcal{U}[-50,20]$ \\
\enddata
\tablecomments{$\mathcal{U}[X,Y]$ indicates a uniform prior with limits $X$ and $Y$, and $\mathcal{N}[X,Y]$ indicates a Gaussian prior with mean $X$, and standard deviation $Y$.}
\end{deluxetable}

\begin{deluxetable*}{l|cc}
\tabletypesize{\scriptsize}
\tablewidth{0pt}
\tablecaption{Transit-Fit Parameters. \label{tab:transfit}}
\tablehead{
    \colhead{Parameter} & \multicolumn{2}{c}{$e$,$\omega$ fixed}}
\startdata
    \multicolumn{3}{c}{Transit Fit Parameters} \\
    \hline
    $\rho_{\star}$ (g cm-3) & \multicolumn{2}{c}{$1.1932^{+0.1090}_{-0.0917}$} \\
    $u_{\textrm{TESS}\,1,1}$ & \multicolumn{2}{c}{$0.1249^{+0.0753}_{-0.0673}$} \\
    $u_{\textrm{TESS}\,2,1}$ & \multicolumn{2}{c}{$0.3030^{+0.0844}_{-0.0727}$} \\
    $u_{\textrm{CHEOPS}\,1,1}$ & \multicolumn{2}{c}{$0.5765^{+0.0433}_{-0.0376}$} \\
    $u_{\textrm{CHEOPS}\,2,1}$ & \multicolumn{2}{c}{$0.2013^{+0.0446}_{-0.0397}$} \\
    $S$ & \multicolumn{2}{c}{$-10.4924^{+0.0573}_{-0.0486}$} \\
    $Q$ & \multicolumn{2}{c}{$1.5277^{+0.0968}_{-0.0787}$} \\
    $\omega_{GP}$ & \multicolumn{2}{c}{$0.8453^{+0.0118}_{-0.0103}$} \\
    \hline
    & planet\,b & planet\,c\\
    \hline
    $T_0$ (TJD) & $1791.0538^{+0.0031}_{-0.0023}$& $1798.0657^{+0.0011}_{-0.0009}$ \\
    $R_P/R_{\star}$ & $0.0438^{+0.0015}_{-0.0011}$ & $0.0434^{+0.0007}_{-0.0006}$ \\
    $b$ & $0.9451^{+0.0061}_{-0.0056}$ & $0.8057^{+0.0152}_{-0.0138}$ \\
    \hline
    \multicolumn{3}{c}{Derived Parameters}\\
    \hline
    & planet\,b & planet\,c\\
    \hline
    $P$ (days)$^1$ & $5.690476^{+0.000004}_{0.000004}$ & $8.538878^{+0.000006}_{0.000005}$\\
    $R_p$ ($R_\oplus$)\,$^2$ & $4.8704^{+0.1622}_{-0.1231}$ & $4.8326^{+0.0736}_{-0.0632}$\\
    $a/R_{\star}$ & $12.6318^{+0.4139}_{-0.2802}$ & $16.5649^{+0.5437}_{-0.3637}$\\
    $i$ (deg)\,$^2$ & $85.6909^{+0.1550}_{-0.0795}$ & $87.2000^{+0.1034}_{-0.0545}$\\
    $a$ (AU)\,$^2$ & $0.0599^{+0.0007}_{-0.0004}$ & $0.0786^{+0.0005}_{-0.0003}$\\
\enddata
\tablenotetext{1}{Average orbital period is derived from a linear fit to the individual transit times.}
\tablenotetext{2}{Values are derived using the $R_*$ value from Table~\ref{tab:prop}}

\end{deluxetable*}

\subsection{Transit Timing Variations}\label{sec:ttv}
The transit timing variations are small within any individual \TESS\ sector and can adequately be fit with a linear ephemeris.
It was only when we combined multiple \TESS\ sectors together, along with the first \CHEOPS\ observations of planet~c -- which revealed a transit 52.49 minutes earlier than predicted by the \TESS\ ephemeris -- that we recognized the presence of significant TTVs in the system.

\begin{deluxetable}{l|l|c|c|c}
\tablewidth{0pt}
\tablecaption{Transit times. \label{tab:transtimes}}
\tablehead{\colhead{planet}& \colhead{t\_N} & \colhead{Time (BTJD)} & \colhead{O-C$^1$ (mins)} & \colhead{Instrument}}
\startdata
        & t$_0$ & 1791.0538 & -74.93 & TESS\\
        & t$_1$ & 1796.7289 & -97.04 & TESS\\
        & t$_3$ & 1808.1266 & -72.91 & TESS\\
        & t$_4$ & 1813.8085 & -85.36 & TESS\\
        & t$_{192}$ & 2883.7340 & 81.63 & TESS\\
        & t$_{193}$ & 2889.4203 & 75.66 & TESS\\
        & t$_{194}$ & 2895.1182 & 86.34 & TESS\\
    planet\,b & t$_{195}$ & 2900.8190 & 101.29 & TESS\\
        & t$_{196}$ & 2906.5131 & 106.46 & TESS\\
        & t$_{252}$ & 3225.1790 & 105.36 & CHEOPS\\
        & t$_{254}$ & 3236.5457 & 84.80 & CHEOPS\\
        & t$_{320}$ & 3612.0110 & -68.00 & TESS\\
        & t$_{321}$ & 3617.6994 & -71.03 & TESS\\
        & t$_{323}$ & 3629.0919 & -54.41 & TESS\\
        & t$_{324}$ & 3634.7904 & -42.85 & TESS\\\hline
        & t$_0$ & 1798.0657 & 21.52 & TESS\\
        & t$_1$ & 1806.5997 & 14.51 & TESS\\
        & t$_{127}$ & 2882.4967 & 12.25 & TESS\\
        & t$_{128}$ & 2891.0234 & -5.35 & TESS\\
        & t$_{129}$ & 2899.5498 & -23.27 & TESS\\
    planet\,c & t$_{130}$ & 2908.0797 & -36.20 & TESS\\
        & t$_{167}$ & 3223.8590 & -265.52 & CHEOPS\\
        & t$_{176}$ & 3300.7982 & -136.82 & CHEOPS\\
        & t$_{213}$ & 3616.8824 & 72.96 & TESS\\
        & t$_{214}$ & 3625.4160 & 65.39 & TESS\\
        & t$_{215}$ & 3633.9592 & 71.64 & TESS\\\enddata
\tablenotetext{1}{minutes early (-), late (+) relative to linear ephemerides.}

\end{deluxetable}

\begin{figure*}
    \centering
    \includegraphics[width=\textwidth]{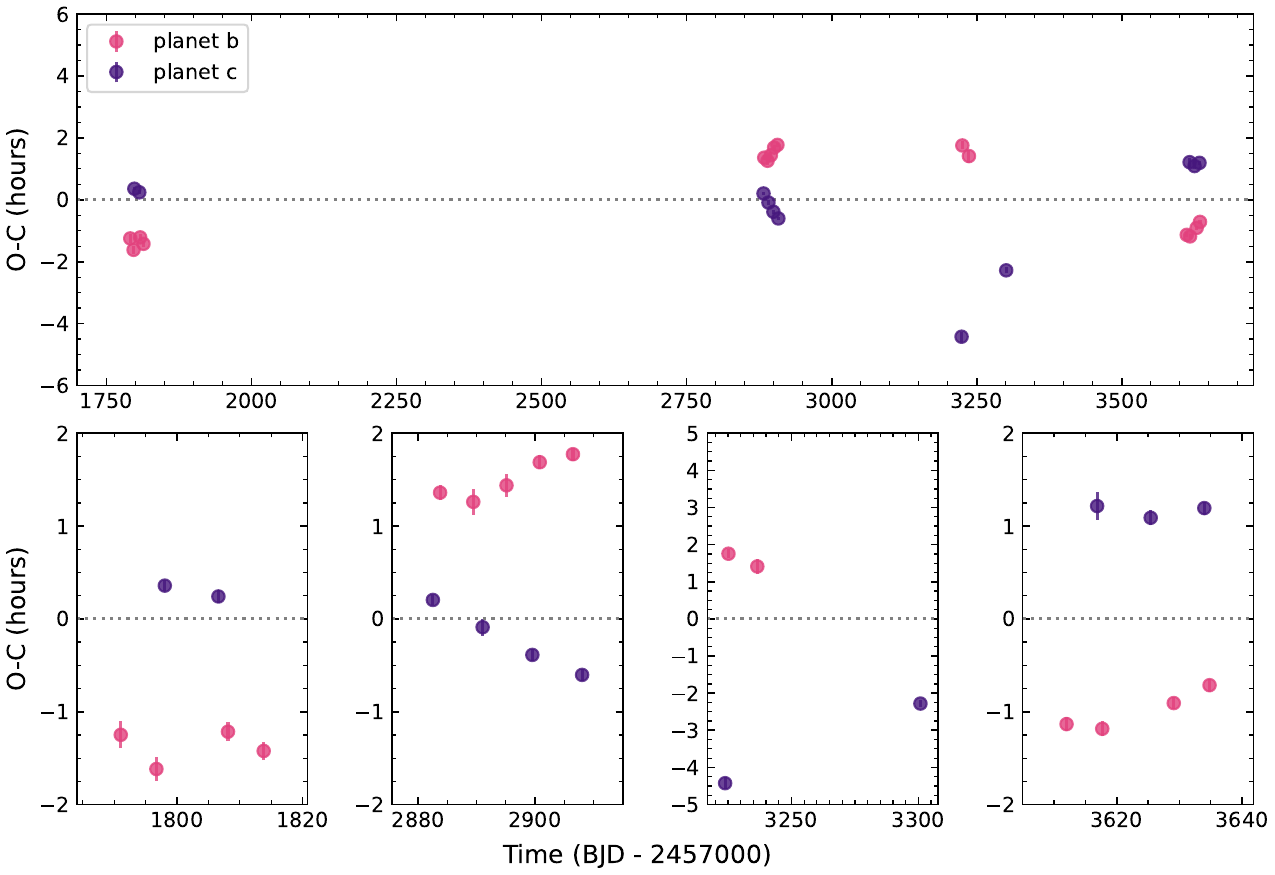}
    \caption{Observed minus calculated transit times for \planetb\ (pink) and \planetc\ (purple) for \added{three} \TESS\ sectors and four \CHEOPS\ visits. The expected transit times are calculated based on fitting a linear period to the observed transit times. There is noted anti-correlation for the TTVs in the second \TESS\ sector (second panel, bottom row). The uncertainties on transit times are smaller than the point size for most transits.}
    \label{fig:ttvs}
\end{figure*}

In our transit fitting, we determined individual transit times and computed the average orbital period by fitting a line to the fits. These transit times can be found in Table~\ref{tab:transtimes}. Figure~\ref{fig:ttvs} shows the O-C diagram.
The TTV curve is not yet well enough sampled to determine parameters, such as planet masses.

\subsection{False positive probability/validation}

We verify that the signals detected in the lightcurve are astrophysical and caused by planets, not binary or background stars, several different ways: statistical validation and through anticorrelated TTVs.

\subsubsection{Statistical Validation}
We use \texttt{TRICERATOPS}\footnote{\url{https://github.com/stevengiacalone/triceratops}} \citep{giacalone2021} to calculate false positive probabilities for both planets. \texttt{TRICERATOPS} is a tool that calculates the probabilities of a range of transit-producing scenarios. To calculate these probabilities, we use the phase-folded data (where each transit is centered at 0) as well as our K-band contrast curve and RV measurements. We additionally use \texttt{molusc}\footnote{\url{https://github.com/woodml/MOLUSC}} to generate binary star priors \citep{wood2021}. For \planetc, the total false positive probability (FPP) is $0.001\pm5.9\times10^{-5}$ and the nearby false positive probability (NFPP) is $0.00\pm0.001$, both of which are well below the thresholds of FPP $<$ 0.015 and NFPP $< 10^{-3}$ that \citet{giacalone2021} suggests for a statistically validated planet. 

The inner planet \planetb, however, has a much more V-shaped transit and in a vacuum does not pass the threshold to be a statistically validated planet. Although the most likely scenario identified by \texttt{TRICERATOPS} is still a planet, the calculation yields significant probabilities for other scenarios, including eclipsing binaries on target or in the background. We can rule out on-target eclipsing binaries because there are no large (km$\,$s$^{-1}$ level) radial velocity variations in our TRES and McDonald spectra. We can also constrain scenarios involving resolved background stars that happen to be eclipsing binaries using our \CHEOPS\ data. We compare transit depths of the \CHEOPS\ light curves for different aperture sizes and find no difference in transit depth between the smallest aperture size and our preferred aperture. Nevertheless, we are not able to rule out all false positive scenarios with this analysis. The probabilities for each of these remaining scenarios are listed in Table~\ref{tab:fp}.

\begin{deluxetable*}{ccccccccccccc}
    \centering
    \tabletypesize{\scriptsize}
    \tablewidth{0pt}
    \tablehead{\colhead{ID} & \colhead{scenario} & \colhead{$M_*$} & \colhead{$R_*$} & \colhead{$ P_{orb}$} & \colhead{inc} & \colhead{b} & \colhead{ecc} & \colhead{w} & \colhead{$R_p$} & \colhead{$M_{EB}$} & \colhead{$R_{EB}$} & \colhead{prob}}
    \startdata
    \hline
        384984325 & TP & 1.03 & 0.95 & 5.69 & 86.92 & 0.68 & 0.30 & 173.65 & 6.35 & 0.00 & 0.00 & 0.32 \\
        384984325 & SEB & 0.24 & 0.26 & 5.69 & 89.44 & 0.35 & 0.01 & 281.15 & 0.00 & 0.10 & 0.12 & 0.10 \\
        384984325 & DTP & 1.03 & 0.95 & 5.69 & 86.82 & 0.63 & 0.38 & 171.07 & 13.86 & 0.00 & 0.00 & 0.12 \\
        384984325 & DEBx2P & 1.03 & 0.95 & 11.38 & 86.55 & 1.60 & 0.15 & 167.09 & 0.00 & 0.98 & 0.95 & 0.10 \\
        384984325 & BEB & 0.88 & 0.87 & 5.69 & 87.88 & 0.32 & 0.74 & 356.72 & 0.00 & 0.74 & 0.77 & 0.04 \\
        384984325 & BEBx2P & 0.67 & 0.60 & 11.38 & 86.58 & 1.09 & 0.74 & 356.97 & 0.00 & 0.66 & 0.60 & 0.01 \\
        384984347 & NEB & 1.00 & 0.61 & 5.69 & 86.75 & 1.37 & 0.27 & 346.06 & 0.00 & 0.37 & 0.38 & 0.25 \\
        384984347 & NEBx2P & 1.00 & 0.61 & 11.38 & 85.95 & 0.95 & 0.71 & 63.94 & 0.00 & 0.99 & 0.61 & 0.05 \\
    \enddata
    \caption{Likely false positives for \planetb\ as calculated by \texttt{TRICERATOPS}. Scenarios with FPP $< 0.01$ are excluded.}
    \label{tab:fp}
\end{deluxetable*}

\subsubsection{Anticorrelated TTVs}
Since our statistical validation was unable to confidently validate the inner planet on its own, we take another approach to demonstrate that both signals in the TOI-6109 system are genuine exoplanets. To do this, we take advantage of the detection of anticorrelated TTVs in the \TESS\ and \CHEOPS\ observations. Over the years, anticorrelation has been used frequently as a way of confidently confirming the existence of near-resonant planetary systems \citep[e.g.,][]{steffen2012}. The argument is that in order for the TTVs to be anti-correlated, they must originate from the same planetary system, and in order for that system to remain stable, their masses must be in the planetary regime.

Transit timing variations are apparent throughout all observations (Figure~\ref{fig:ttvs}). In the Sector 58 \TESS\ data the TTVs from either planet are anticorrelated. This is to be expected of two planets mutually interacting near a first-order mean motion resonance and are unlikely to be produced by photometric noise \citep{steffen2012}.

To test the likelihood that the transit timing variations are indeed anticorrelated, we fit for the parameter $\Xi_{max}$, which is an anticorrelation measurement for the transit timing variations, as described in \citet{steffen2012}. 

We fit the transit timing residuals with the equation
\begin{equation}\label{eq:f}
    f = A \sin{\frac{2\pi t}{P_i}} + B \cos{\frac{2\pi t}{P_i}} + C
\end{equation}

via a least-squares algorithm, where $A,B,C$ are model parameters and $P_i$ is a range of test timescales ($P_i = [1, 1500]$ days. The statistic is calculated as

\begin{equation}\label{eq:xi}
    \Xi = - \left( \frac{A_1 A_2}{\sigma_{A_1}\sigma_{A_2}} + 
                    \frac{B_1 B_2}{\sigma_{B_1}\sigma_{B_2}} \right)
\end{equation}  

\begin{figure}
    \centering
    \includegraphics[width=\columnwidth]{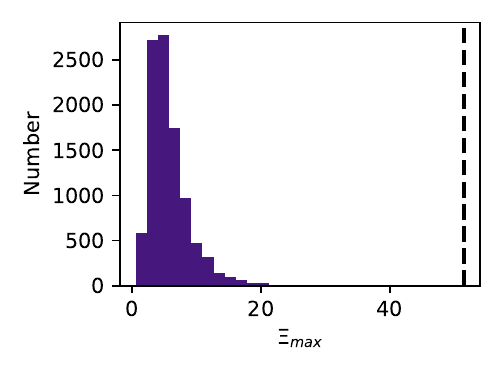}
    \caption{$\Xi_{max}$ value for 10,000 Monte Carlo runs of Equation~\ref{eq:xi}, shuffling our TTV residuals each time. Our real data have a \added{$\Xi_{max}$=51.43 (dashed line), which is 13\,$\sigma$ larger} than the median of the shuffled runs. There is extremely little chance the TTVs observed could be from photometric noise.}
    \label{fig:xi}
\end{figure}

Our maximum $\Xi$ value is \added{$\Xi_{max} = 51.43$ at $P_i = 792$ days.} To test the likelihood that the transit timing variations are anticorrelated, we shuffle the transit timing residuals, refit the orbital period, calculate the new transit time residuals, and fit Equation~\ref{eq:f} and Equation~\ref{eq:xi} for the range of test timescales. We perform this test 10,000 times. We plot the distribution of maximum values, $\Xi_{max}$ in Figure~\ref{fig:xi}. None of the tests measured a $\Xi_{max}$ value as high as the original data, meaning it is statistically unlikely that the TTVs are due to photometric noise.

\subsection{Mean Motion Resonance}
Given the detection of TTVs in the TOI-6109 system, we performed simulations to study the dynamical state of the system. First, we test whether the orbital configuration is stable and whether it is likely the planets are in a true mean motion resonance (MMR). To this end, we conduct a set of 1,000 $N$-body simulations of the system using the \texttt{rebound} code \citep{ReinLiu2012}, with initial conditions randomly drawn from the posteriors in Table \ref{tab:transfit}. Our simulations made use of \texttt{whfast}, a symplectic Wisdom--Holman integrator \citep{WisdomHolman1991, ReinTamayo2015}. Each simulation is run to $10^5$ orbits of the innermost planet, with a timestep of $0.05$ times that orbital period. About 2.7\% of these configurations are unstable. Of the remainder, we find that 29\% of these cases are absolutely in the 3:2 MMR (in pure resonance). A further 15\% are in quasi-resonant states or ``nodding'' between resonances \citep[i.e.,][]{Ketchum2013, Khain2020}. The resonant angle most commonly seen to librate is $3\lambda' - 2\lambda - \varpi'$. It is therefore ambiguous whether these planets are in true MMR, but true resonance remains a strong possibility.

\subsection{Injection/Recovery Analysis}
\begin{figure}
    \centering
    \includegraphics[width=\linewidth]{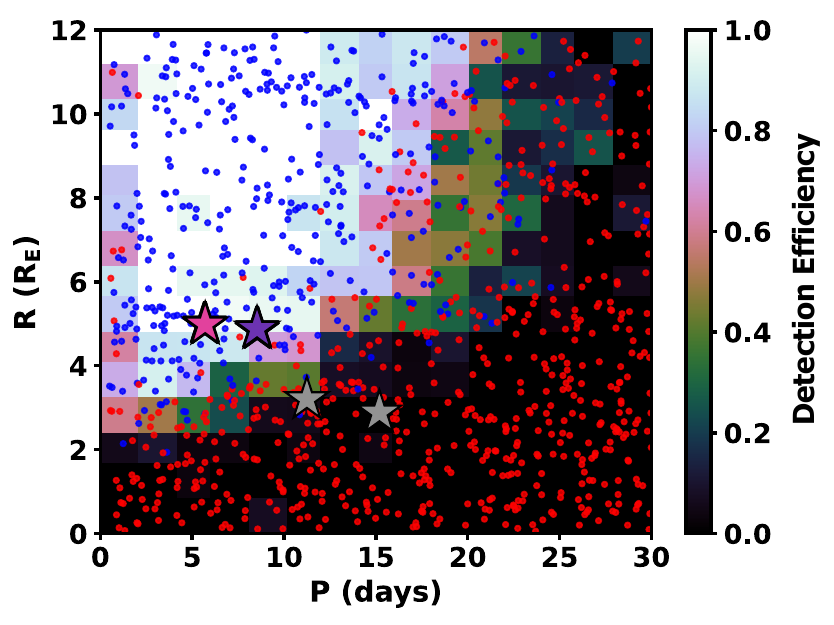}
    \caption{Period-radius diagram of injected and recovered targets. Blue points are those that were recovered by our pipeline. The pink star is \planetb\ and the purple star \planetc. The unconfirmed planet candidates are shown in gray.}
    \label{fig:injections}
\end{figure}

We run injection/recovery tests \added{on the TESS sector 18 and 58 data} for this star to characterize the detection sensitivity of our detection pipeline. If these targets lie in a low-completeness area of parameter space they are more likely to be false alarms.
We injected 5120 targets into our detection pipeline (described in Section~\ref{sec:transits}) pulled from a log-uniform distribution in period and radius. Figure~\ref{fig:injections} shows the results of this process. Both planets are within a part of parameter space with high detection completeness.
The other two candidates, at 11.5~days and 15.7~days, lie in the low-threshold regime and therefore are difficult to confirm with \TESS\ data alone. 

While both planets cannot be individually statistically validated, we combine multiple lines of evidence to confirm these planets. Both planets are in a high area of recovery for our detection pipeline. Planet~c can be validated with \texttt{TRICERATOPS}, and the most likely scenario remains True Planet for planet~b.
The presence of anti-correlated TTVs mean that the bodies have to be dynamically interacting in the same system.
Dynamical simulations show that these planets are in a stable, near mean motion resonant configuration, which means that they must be planetary mass bodies.

\section{Discussion}\label{sec:discussion}

\added{\subsection{Are all young planets in MMR?}\label{sec:young-mmr}
There is a growing sample of planets with ages less than 100\,Myr. \planetb\ and \planetc\ are well-representative of this sample; they are at short orbital period, have large--likely inflated--radii, and the system is in near mean motion resonance. While this is similar to the young planets we have detected, is it representative of young planets overall?

\begin{figure}
    \centering
    \includegraphics[width=\columnwidth]{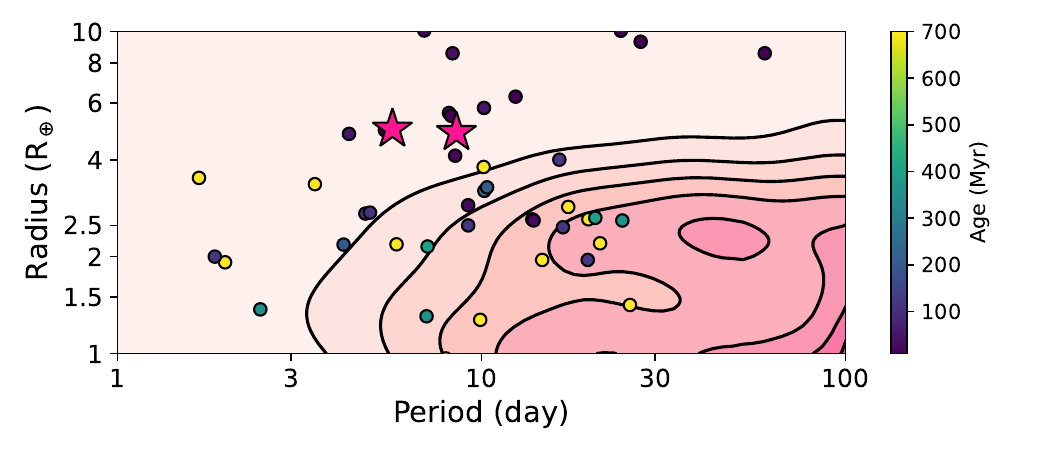}
    \caption{86 young planets ($<1\,$Gyr) plotted via their orbital period and radius, colored by their age. The contours below are planet occurrence contours from \citet{dattilo2023}. The population of young planets is not a representative sample of planets compared to the older, intrinsic population of the galaxy, due to various detection biases.}
    \label{fig:young-occ}
\end{figure}

Young planets are overall larger than their older counterparts (e.g., those from Kepler, where planets between the size of Earth and Neptune make up the bulk of planet detections). We plot the sample of confirmed planets with ages less than 1\,Gyr and orbital periods less than 100 days from the Exoplanet Archive\footnote{\url{https://exoplanetarchive.ipac.caltech.edu/}, accessed 2025-May-08} on top of Kepler occurrence rate contours from \citet{dattilo2023} in Figure~\ref{fig:young-occ} for a comparison of the young planet sample and the several-Gyr-old planet occurrence. The authors hesitate to declare all young planets to be larger; there are a host of observational biases that make this appear true, rather than it reflecting the intrinsic population of young planets in the galaxy. 

These planets, and likely the outer two planet candidates we do not confirm, are in near mean motion resonance. This is  very common for young planetary systems, as seen in \citet{dai2024}, who found that $70\pm15$\,\% of young ($< 100\,$Myr) systems are in mean motion resonance versus $15\pm2$\,\% of mature ($> 1\,$Gyr) systems.

It is possible that the young planet sample is being biased towards multi-planet systems in near-resonant configurations, particularly if TTVs are used to confirm secondary planets in the system, as we have done here. This can lead us to draw conclusions about planetary formation that are not accurate. This highlights the need to do demographic studies on planet samples that are detected and vetted in a uniform manner rather than on samples of confirmed planets; the demographic studies brought up in the introduction \citep{christiansen2023, vach2024, fernandes2025} all build their own pipelines for planet detection and use those resulting planet samples for analysis.}

\subsection{Atmospheric Evolution}\label{sec:atmos_evol}
\begin{figure}
    \centering
    \includegraphics[width=\columnwidth]{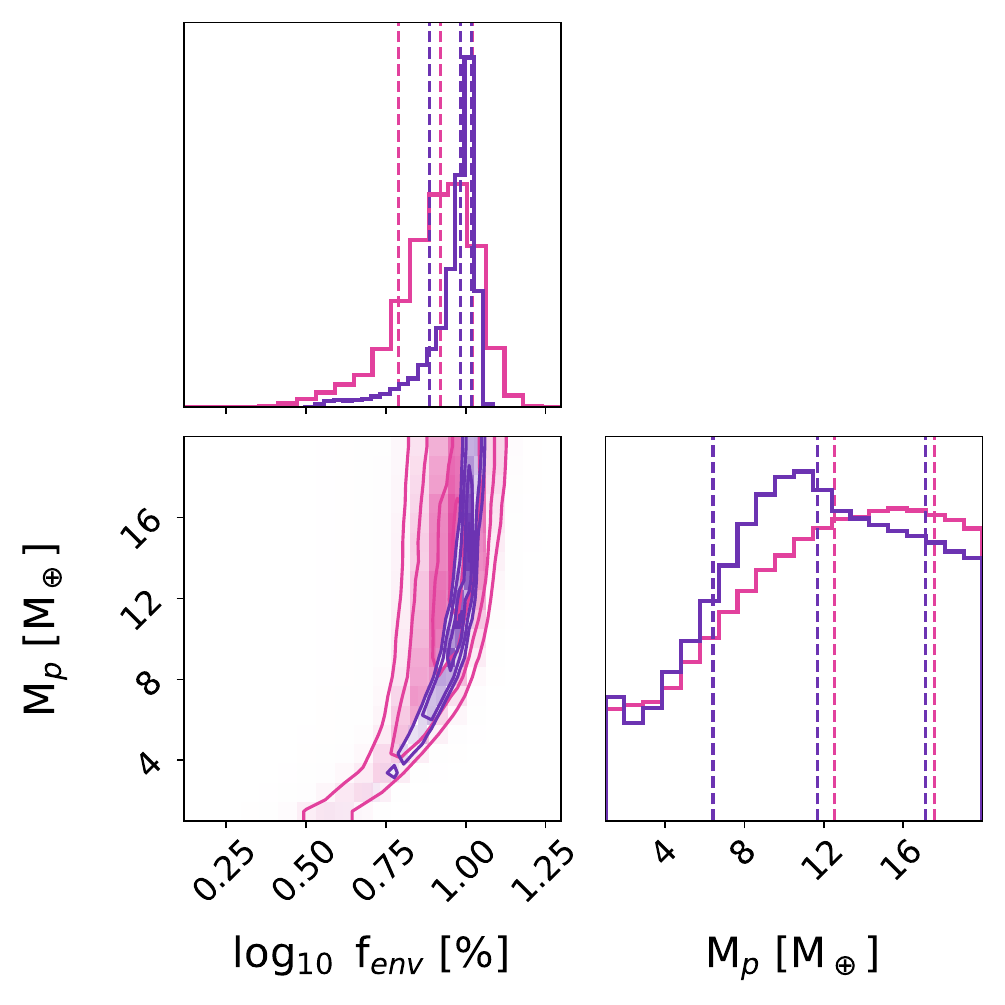}
    \caption{Posterior composition distributions for \planetb\ (pink) and \planetc\ (purple) from \texttt{smint}. The possible planet masses are 4--18~M$_\oplus$ for both planets. The envelope mass fractions are 7.94\% for \planetb\ and 9.55\% for \planetc.}
    \label{fig:smint}
\end{figure}

Because the radii of sub-Neptune and Neptune-sized planets are mostly independent of their mass, radii can be used as a proxy for atmospheric mass fraction \citep{lopez2014}. We use \texttt{smint}\footnote{\url{https://github.com/cpiaulet/smint}} \citep[Structure Model INTerpolator][]{piaulet2021} to constrain potential H/He envelope mass fractions for a range of planet masses\footnote{We assume these planets are rocky cores with large H/He envelopes. Due to their size ($>4$\Re), they are not likely to be steam worlds \citep{aguichine2021}.}. \texttt{smint} interpolates over a grid of envelope mass fractions ($f_{env}$), mass, age, and insolation to fit an observed planet radius. We run a MCMC to best-fit these parameters for our planet radii, using a uniform prior on $f_{env}$ over the entire grid range of $f_{env} = [0.01, 1.0]$, a Gaussian prior based on our measured values for insolation, and wide prior on mass, $15\pm10$M$_\oplus$ for both individual planets. These model grids are based on \citet{lopez2014} and are limited to ages of [0.1-10]~Gyr. We set the age to be 0.1~Gyr, which is higher than the measured age of the system. This effectively places a lower limit on the envelope mass fraction. Figure~\ref{fig:smint} shows the posterior distributions for a 1~solar metallicity model with the age constrained to 0.1~Gyr. We can place a lower limit on the current envelope mass fraction to be 7.94\% for \planetb\ and 9.55\% for \planetc.

The distribution of core masses ranges from 5\,M$_\oplus$ to 18\,M$_\oplus$.
These planets are young enough to still be undergoing atmospheric mass loss, and they will ultimately cool and contract to a smaller radius after a billion years. Based on the sub-Neptune mass distribution from \citet{polanski2024}, these planets will likely end up as large sub-Neptunes, 3.5--2.5\Re.

\subsection{Additional Planet Candidates}
During our transit searches, we identified two additional transit signals. To confirm these signals, we attempted to observe additional transits with \CHEOPS; however, all four observations were unsuccessful in detecting transits. Given the time gap between the last \TESS\ observations and the \CHEOPS\ follow-up, it is possible that the predicted ephemeris had drifted or that significant TTVs caused the transits to be missed. \added{Due to scattered light in the second-half of the Sector 85 data, we were unable to observe consecutive transits of either additional candidate signal.} We reserve judgment of the validity of these signals as planets for a future paper.

\subsection{Follow-up potential}
We calculate the transmission spectroscopy metric \citep[TSM,][]{Kempton2018} for TOI-6109~b and c to estimate their potential for atmospheric characterization with JWST. Using the definition in the original paper, TOI-6109~b and c have TSM values of 140 and 118, respectively, readily exceeding the recommended threshold for followup in their radius regime and indicating atmospheres with high potential for transmission spectroscopy. This formulation underestimates the planets' potential since it relies on the \citet{chen2017} mass-radius relationship, which is likely to be an overestimate for these young planets. Using the masses estimated in Section \ref{sec:atmos_evol} instead yields TSM values of 247 and 218, indicating superlative potential for JWST transmission spectroscopy. For context, the TOI-6109 planets have some of the highest TSM values than other exoplanets with R$_p<5$R$_\oplus$ and host stars younger than 500\,Myr.

While these planets are young enough to still be undergoing atmospheric mass loss, mass loss observations of the system would be challenging. The system is $148.5\pm0.4$~pc away \citep{gaiadr3}, which is too far to measure Lyman-$\alpha$ emission from the planets, due to interstellar absorption. The metastable Helium lines are also not likely to be observed due to its stellar type; these observations have only been successfully observed for K-type stars.

\section{Summary \& Conclusion}\label{sec:conc}
Using space- and ground-based data, we confirm two planets around the star TOI-6109. We find that:
\begin{enumerate}
    \item The star \thisstar\ is a member of the Alpha Persei cluster and is 75$\pm$5\,\added{Myr} old. We verify its membership by its galactic position, kinematics, rotation, and lithium abundance.
    \item We confirm two planets using \TESS\ and \CHEOPS\ data:
    \begin{enumerate}
        \item \planetb\ is a \radb\ \Re\ planet on a \perb\ day orbit.
        \item \planetc\ is a \radc\ \Re\ planet on a \perc\ day orbit.
    \end{enumerate}
    \item This system is in a near 3:2 mean motion resonance and report on the system's transit timing variations.
\end{enumerate}

Young planetary systems like TOI-6109 are critical for understanding the processes of planet formation and early evolution. The system's youth and near-resonant state make it a prime candidate for further investigation. Continued transit monitoring of TOI-6109 will refine the TTV-derived masses and enable detailed studies of its dynamical interactions. These efforts will contribute to our broader understanding of how planets form, migrate, and settle into their final configurations.

\begin{acknowledgments}
    We thank the anonymous referee for their helpful comments that improved this manuscript.
    This paper includes data collected by the \TESS\ mission, which are publicly available from the Mikulski Archive for Space Telescopes (MAST). Funding for the \TESS\ mission is provided by NASA’s Science Mission Directorate.

    CHEOPS is an ESA mission in partnership with Switzerland with important contributions to the payload and the ground segment from Austria, Belgium, France, Germany, Hungary, Italy, Portugal, Spain, Sweden, and the United Kingdom. The CHEOPS Consortium would like to gratefully acknowledge the support received by all the agencies, offices, universities, and industries involved. Their flexibility and willingness to explore new approaches were essential to the success of this mission. CHEOPS data analysed in this article will be made available in the CHEOPS mission archive (\url{https://cheops.unige.ch/archive_browser/}).
    
    This research has made use of the Exoplanet Follow-up Observation Program (ExoFOP; DOI: 10.26134/ExoFOP5) website, which is operated by the California Institute of Technology, under contract with the National Aeronautics and Space Administration under the Exoplanet Exploration Program.
    
    Some of the observations in this paper made use of the High-Resolution Imaging instrument ‘Alopeke and were obtained under Gemini LLP Proposal Number: GN/S-2021A-LP-105. ‘Alopeke was funded by the NASA Exoplanet Exploration Program and built at the NASA Ames Research Center by Steve B. Howell, Nic Scott, Elliott P. Horch, and Emmett Quigley. Alopeke was mounted on the Gemini North telescope of the international Gemini Observatory, a program of NSF’s OIR Lab, which is managed by the Association of Universities for Research in Astronomy (AURA) under a cooperative agreement with the National Science Foundation. on behalf of the Gemini partnership: the National Science Foundation (United States), National Research Council (Canada), Agencia Nacional de Investigación y Desarrollo (Chile), Ministerio de Ciencia, Tecnología e Innovación (Argentina), Ministério da Ciência, Tecnologia, Inovações e Comunicações (Brazil), and Korea Astronomy and Space Science Institute (Republic of Korea).
    
    This work makes use of observations from the LCOGT network. Part of the LCOGT telescope time was granted by NOIRLab through the Mid-Scale Innovations Program (MSIP). MSIP is funded by NSF.

    This research was carried out in part at the Jet Propulsion Laboratory, California Institute of Technology, under a contract with the National Aeronautics and Space Administration (80NM0018D0004).
    
    A.D. gratefully acknowledges support from the Heising-Simons Foundation through grant 2021-3197.
    R.K. acknowledges funding from the Dunlap Institute, which is funded through an endowment established by the David Dunlap family and the University of Toronto.
    A.M. acknowledges funding from a UKRI Future Leader Fellowship, grant number MR/X033244/1 and a UK Science and Technology Facilities Council (STFC) small grant ST/Y002334/1.
    JGM acknowledges support from the Heising Simons Foundation through a 51 Pegasi B Fellowship, and from the Pappalardo family through the MIT Pappalardo Fellowship in Physics. The Tierras Observatory is supported by the National Science Foundation under Award No. AST-2308043.
    
    We recognize and acknowledge the very significant cultural role and reverence that the summit of Mauna Kea has always had within the indigenous Hawaiian community. We are grateful to have the opportunity to conduct observations from this sacred mountain.

    All \TESS data used in this paper can be found in MAST: \dataset[doi:10.17909/0smj-7050]{https://doi.org/10.17909/0smj-7050}. The raw and detrended photometric time-series data for all eight CHEOPS observations are available in electronic form at the CDS via anonymous ftp to cdsarc.u-strasbg.fr (130.79.128.5) or via \url{http://cdsweb.u-strasbg.fr/cgi-bin/qcat?J/A+A/}.

\end{acknowledgments}

\facilities{\TESS, Mikulski Archive for Space Telescopes (MAST), Gaia, LCGOT}

\software{\textit{Astropy} \citep{astropy}, \textit{numpy} \citep{numpy}, \textit{pandas} \citep{pandas}, \textit{scipy} \citep{scipy}, \textit{matplotlib} \citep{matplotlib}, \textit{batman} \citep{kreidberg2015}, \textit{emcee} \citep{emcee}, \textit{dynesty} \citep{speagle2020}, \textit{juliet} \citep{espinoza2019}, AstroImageJ \citep{collins2017}, TAPIR \citep{jensen2013} \textit{gyro-interp}\citep{bouma2023}, \textit{EAGLES}\citep{jefferies2023}, \textit{galpy}\citep{bovy2015}}

\bibliography{sample631}{}
\bibliographystyle{aasjournal}

\end{document}